\documentclass[twocolumn,tighten]{aastex63}
\usepackage{graphicx}
\usepackage{gensymb}
\usepackage{xcolor}

\shorttitle{Late-Time Emissions from SN 1996cr}
\shortauthors{Patnaude et al.}

\begin{document}

\title{Late-Time Optical and X-ray Emission Evolution of the Oxygen-Rich SN~1996cr}

\author[0000-0002-7507-8115]{Daniel Patnaude}
\affil{Smithsonian Astrophysical Observatory, 60 Garden Street, Cambridge, MA 02138 USA}

\author[00000-0002-4471-9960]{Kathryn E.\ Weil}
\affil{Smithsonian Astrophysical Observatory, 60 Garden Street, Cambridge, MA 02138 USA}
\affil{6127 Wilder Lab, Department of Physics and Astronomy, Dartmouth
                 College, Hanover, NH 03755 USA}
\affil{Department of Physics and Astronomy, Purdue University, 525 Northwestern Avenue, West Lafayette, IN 47907 USA}
                 
\author[0000-0003-3829-2056]{Robert A.\ Fesen}
\affil{6127 Wilder Lab, Department of Physics and Astronomy, Dartmouth
                 College, Hanover, NH 03755 USA}

\author[0000-0002-0763-3885]{Dan Milisavljevic}
\affil{Department of Physics and Astronomy, Purdue University, 525 Northwestern Avenue, West Lafayette, IN 47907 USA}
\affiliation{Integrative Data Science Initiative, Purdue University, West Lafayette, IN 47907, USA}

\author[0000-0002-0765-0511]{Ralph P.\ Kraft}
\affil{Smithsonian Astrophysical Observatory, 60 Garden Street, Cambridge, MA 02138 USA}
        

\begin{abstract}

When the ejecta of supernovae interact with the progenitor star's circumstellar environment, a strong shock is driven back into the ejecta, causing the material to become bright optically and in X-rays. Most notably, as the shock traverses the H-rich envelope, it begins to interact with metal rich material. Thus, continued monitoring of bright and nearby supernovae provides valuable clues about both the progenitor structure and its pre-supernova evolution. Here we present late-time, multi-epoch optical and {\sl Chandra} X-ray spectra of the core-collapse supernova SN 1996cr. Magellan IMACS optical spectra taken in July 2017 and August 2021 show a very different spectrum from that seen in 2006 with broad, double-peaked optical emission lines of oxygen, argon, and sulfur with expansion velocities of $\pm 4500$ km s$^{-1}$. Red-shifted emission components are considerably fainter compared to the blue-shifted components, presumably due to internal extinction from dust in the supernova ejecta. Broad $\pm 2400$ km s$^{-1}$ H$\alpha$ is also seen which we infer is shocked progenitor pre-SN mass-loss, H-rich material. \textit{Chandra} data indicate a slow but steady decline in overall X-ray luminosity, suggesting that the forward shock has broken through any circumstellar shell or torus which is inferred from prior deep \textit{Chandra} ACIS-S/HETG observations. The X-ray properties are consistent with what is expected from a shock breaking out into a lower density environment. Though originally identified as a SN IIn, based upon late time optical emission line spectra, we argue that the SN 1996cr progenitor was partially or highly stripped, suggesting a SN IIb/Ib.

\end{abstract}


\section{Introduction}

A supernova's (SN) rapid expansion naturally leads to ever decreasing ejecta column densities, optical opacities, and temperatures. This, in turn, leads to fewer optically thick lines and the eventual emergence of an emission-line dominated nebula at late-times.  Consequently, optical and X-ray studies of SNe in this optically thin nebular phase of evolution can provide valuable insights into the kinematic and elemental properties of both outer and interior supernova ejecta.

Observations of SN emissions many years or decades following maximum light are relatively rare. While optical observations of even relatively bright Type Ia SNe have been limited to $\sim$1500 days post-maximum light \citep{Graham2015,Taub2015,Kerz2017}, late-time radio, optical, and X-ray emissions have been detected from a handful of core-collapse SNe (CCSNe) several decades after outburst \citep{Long1989,Fesen1990,Cowan1994,Eck2002,Soria2008,Mili2012,Dwark2012,FW2020}. 

\begin{figure*}
\centering
\includegraphics[width=0.97\textwidth]{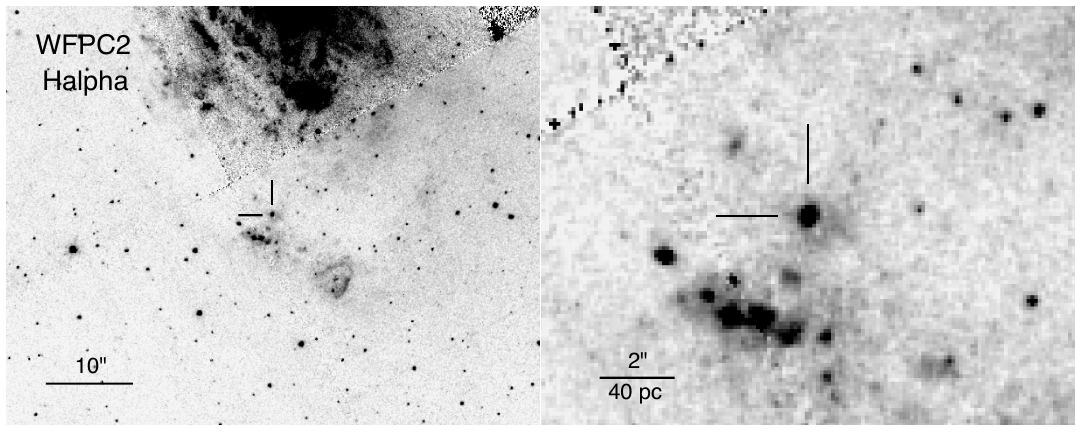}
\caption{{\bf{Left:}} A 1999 HST/WFPC2 F656N H$\alpha$ image of SN~1996cr showing the supernova's location south of the nucleus of the Circinus galaxy along with neighboring H~II regions. {\bf{Right:}} Zoom-in of the same H$\alpha$ image showing the supernova lies coincident with  relatively faint and extended H~II region emission.
}
\label{HST_image}
\end{figure*}
\begin{figure}
    \centering
    \includegraphics[width=0.98\columnwidth]{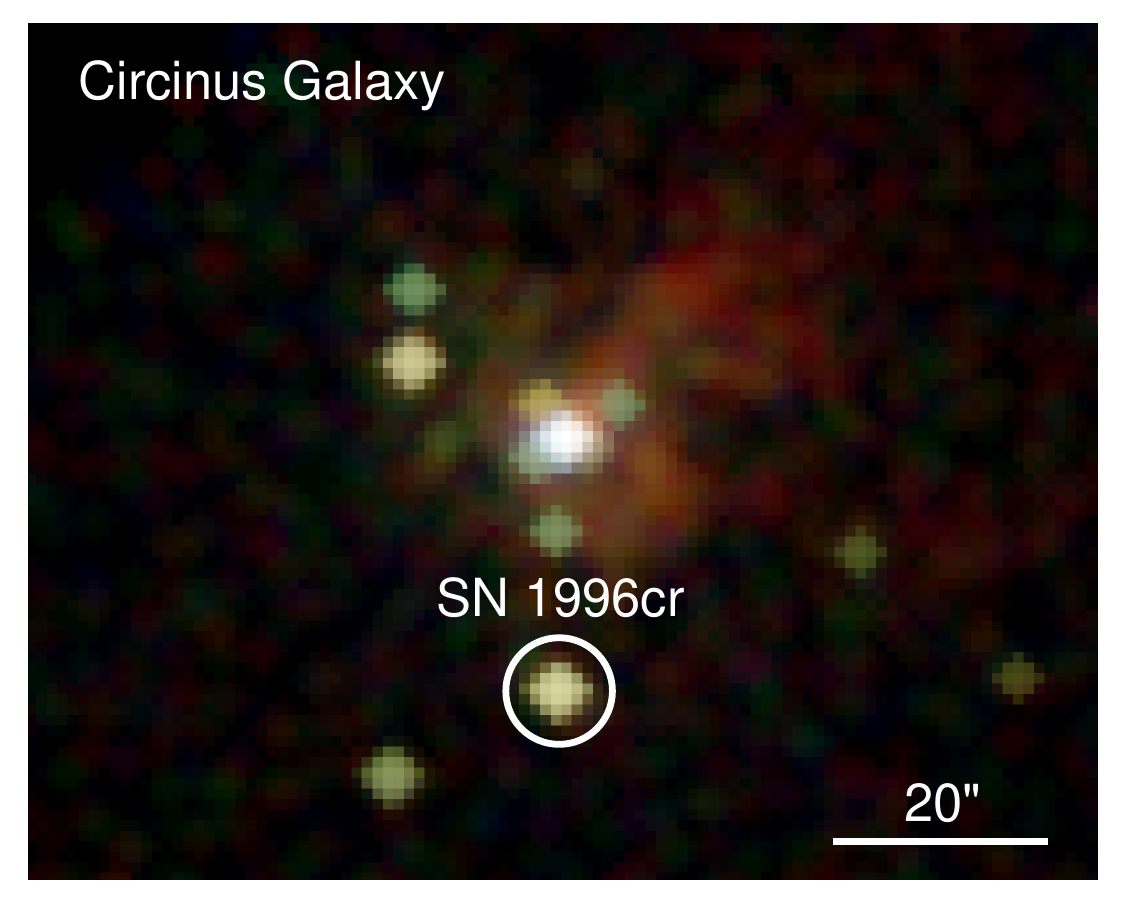}
    \caption{Color {\it Chandra} ACIS-S image of Circinus Galaxy and SN~1996cr obtained 30 November 2018.}
    \label{fig:tcolor}
\end{figure}

The main energy source for late-time CCSN emissions is not believed to be the decay energy of radioactive isotopes such as $^{56}$Ni or $^{44}$Ti as is the case for type Ia SNe \citep[e.g.,][]{Mili2012}. Instead, the usual interpretation for the relatively high luminosities of CCSNe years and even decades after outburst across multiple wavelengths is the interaction of the SN's expanding ejecta with circumstellar material (CSM) \citep{Chevalier1982,CF94,Chugai2004,Smith2017,Patnaude2017,CF2017}.

While other late-time energy sources have been suggested including pulsars, accretion around stellar mass black holes, and magnetars \citep{CF94,Woosley2010,Patnaude2011,Mili2018}, the presence of broad, high-velocity ($>$ 1000 km s$^{-1}$) H$\alpha$ emission in late-time CCSN spectra has been generally viewed as the signature of the interaction of SN ejecta with hydrogen-rich circumstellar medium \citep{Mili2015,Smith2017}. 

The collision of high-velocity SN ejecta with pre-SN progenitor wind or dense CSM in the form of a clumpy shell or disk local to the progenitor star will convert a portion of the ejecta's kinetic energy into radiation due to shock heating of the CSM and strong shock heating of the SN ejecta \textcolor{red}{\citep{CF94,chevalier06}}. This results in both broad metal-rich emission lines from shocked SN ejecta and broad hydrogen line emission from the shocked CSM \citep{Smith2017}.

The amount of progenitor mass loss, the density of the CSM and when it was released prior to the SN explosion determines when the ejecta-CSM interaction occurs and the resulting luminosity of both the shocked CSM and SN ejecta. CCSNe sub-types which exhibit late-time emission due to ejecta-CSM interaction include IIn, II-L, IIb, and Ibn, with type IIn showing the strongest late-time optical, radio, and X-ray emissions due to an estimated mass of surrounding CSM of several solar masses \citep{Fransson2014,Dessart2015}.  

Dust formed in either the SN ejecta or present in the pre-SN mass loss CSM can lead to an optical spectrum where the blueshifted line emissions from the facing hemisphere of ejecta are considerably brighter than that from the higher extinction, rear hemisphere ejecta \citep{Fesen1990,Fesen1993,McCray1993,CF94,Andrews2010,Andrews2011,Mili2012,Bevan2017}. A notable exception is the recently detected SN~1941C where the emission from its redshifted ejecta is much brighter than its blueshifted ejecta \citep{Jones41,FW2020}.

\subsection{CCSNe with Late-Time Emissions}

There is currently only a handful of CCSNe with reported late-time radio, optical, and X-ray emissions more than one or two decades post-max and these are most often SN IIn, SN~II-L, SN~IIb, or SN~Ib subtypes \citep{Weiler2007,Dwark2012,Mili2012,Kundu2019,FW2020,Ram2021}. In these and similar cases, the late-time emission is likely due to significant CSM immediately local to the SN. The subclass of SNe~IIn exhibit the most extreme CSM environments due to very high estimated progenitor mass loss rates of $\sim 10^{-4}$ to 0.1 M$_{\odot}$  yr$^{-1}$\citep{Sala1998,Gal-Yam2009,Kiewe2012,Smith2017}. Early examples of SNe~IIn include SN~1978K, SN~1988Z, SN~1994W, and SN~1995N \citep{Ryder1993,Turatto1993,Fransson2002,Chugai2004}. 

In contrast, the more common CCSNe arising from red or yellow supergiant progenitors observed as SNe II-L and SNe IIb have considerably smaller mass loss rates ($\sim 10^{-5}$  to $10^{-6}$ M$_{\odot}$ yr$^{-1}$) along with possible episodes of much higher mass loss rates hundreds to thousands of years before outburst. Early observed examples of such objects include SN~1970G, SN~1979C, SN~1980K, and SN~1993J. Strong emissions decades after the explosion are not commonly present for the subclass of SNe II-P presumably due to their lower progenitor mass loss rates and thus less dense CSM environment prior to their explosion, although there are exceptions (SN 2004et, \citealt{Long2019}; SN~2007od, \citealt{Andrews2010}; and SN~2017eaw, \citealt{Weil2020a}).

A few 50 to 100 yr old extragalactic supernova remnants (SNRs) not associated with observed historical SNe have been found to be ejecta-dominated, including the luminous O-rich SNR in NGC~4449 \citep{Kirshner1980,Blair1983,Long2012} and B174a in M83 \citep{Blair2015}. In these cases, broad H$\alpha$ emissions are again seen, supporting the CSM--ejecta interaction scenario as the chief late-time energy source in some CCSNe. The most distinct difference between the two is the relative emission lines strength of [\ion{O}{3}]/([\ion{O}{1}]+[\ion{O}{2}]), which is known to increase with age. 

\subsection{Late-Time Emissions from SN 1996cr}

One of the X-ray brightest CCSN showing evidence of strong ejecta--CSM interaction is SN 1996cr in the Circinus galaxy (d $\simeq$ 4.2 Mpc, z=0.001448; \citealt{Freeman1977,Korib2004,Tully2008,Tully2009,Lianou2019}). The SN is located at
$\alpha$ = $14^{\rm h} 13^{\rm m} 10.01^{\rm s}$, $\delta$ = $-65\degr 20' 44.4''$ (J2000), some $23''$ south of the galaxy's center coincident with a faint diffuse H~II region, with a line of much brighter H~II regions $\sim3''$ ($\sim$60 pc) off to the southeast (see Figs.~\ref{HST_image} \& \ref{fig:tcolor}). The Circinus galaxy itself is located in a region of relatively low Galactic absorption with an $A_{\rm V} = 1.5 \pm 0.3$ and an $N_{\rm H} = 3.3 \pm 0.3$ $\times 10^{21}$ cm$^{-2}$  \citep{Freeman1977,Schlegel1998}.

The SN~1996cr remnant was first discovered as a discrete X-ray source (CG X-2) with properties suggestive of a young ultra-luminous SNR \citep{Sambruna2001,Bauer2001}. Analysis of archival data along with follow-up radio, optical and X-ray observations later confirmed it to be a young SNR, possibly associated with SN IIn type event due to strong H$\alpha$ emission. These data indicate a likely explosion date between 28 February 1995 and 16 March 1996 \citep{Bauer2007,Bauer2008}. Radio data showed SN~1996cr with a spectral index of $-0.85$ along with evidence of a spectral flattening by day 3000 \citep{Bauer2008,Meunier2013}. The SN underwent more than a 30 fold increase in X-ray flux between 1997 and 2000 presumably due its interaction with a dense surrounding circumstellar environment \citep{Bauer2008}. 

The first optical spectrum of SN~1996cr, taken in early 2006 when the remnant was already some 10 yr old, revealed a highly reddened (A$_{\rm V} \simeq 6$) object with multiple component emission lines of hydrogen and forbidden oxygen spanning radial velocities of 2000 to 5800 km s$^{-1}$ along with narrower lines of hydrogen, helium, [\ion{N}{2}], [\ion{S}{2}], and [\ion{Fe}{7}] \citep{Bauer2008}. 

The structure of the oxygen emission lines was especially complex. Roughly symmetrical blue and red emission profiles for the [\ion{O}{1}], [\ion{O}{2}], and [\ion{O}{3}] emission lines with  velocities of $+3600$ and $-3250$ km s$^{-1}$ were seen, plus a separate much higher velocity blueshifted component at $-5750$ km s$^{-1}$, along with a broadened central component with a FWHM of $2800$ km s$^{-1}$.

Analyses of various line ratios led \citet{Bauer2008} to suggest SN~1996cr was a Type IIn event where the progenitor exploded in a low density medium through which the ejecta expanded through for the first few years, and then subsequently hit a dense, wind-swept CSM shell. They estimated the dense shell's radius at 0.03 pc with an estimated electron density, $n_{e} \geq 10^{5}$ cm$^{-3}$. Assuming that the observed extinction corrected H$\alpha$ luminosity was from photoionized stellar wind material, \citet{Bauer2008} also estimated a progenitor mass loss rate of $\sim10^{-5}$ M$_{\odot}$ yr$^{-1}$. VLBI measurements obtained in June 2007 (age $\simeq 12$ yr) suggested a forward shock radius of $2.8 \times 10^{17}$ cm (0.09 pc).

Based largely on X-ray data, \cite{Bauer2008} further proposed a scenario where, sometime prior to its explosion, the progenitor star transitioned to a Wolf-Rayet like phase with a fast dense wind. This formed a low density wind-blown cavity which was surrounded by a dense CSM shell of earlier red supergiant (RSG) mass loss material. 

This cavity + shell CSM model was supported by X-ray observations that showed it was undetected at day $\sim$ 1000 but brightening to $L_{x} \sim 4 \times 10^{39}$ erg s$^{-1}$ (0.5-8 keV) after 10 yr, and by hydrodynamic simulations and modeling of high resolution {\sl Chandra} HETG X-ray spectra \citep{Dewey2010}. More recent analyses of SN~1996cr's X-ray emission spectra have concluded that an asymmetrical polar cap geometry plus internal obscuration could reproduce the observed X-ray line profiles \citep{Dewey2011,QV2019}. SN~1996cr's current X-ray emission, although fading, still places it among the most luminous late-time X-ray CCSNe known \citep{Ross2017,QV2019}.

Here we present late-time optical and X-ray spectra of SN~1996cr, which along with recent radio and X-ray analyses \citep{Meunier2013,QV2019}, help to formulate a more complete picture of this young CCSN remnant. We also place it in context with new examples of SNe IIb/Ib/Ic explosions from stripped-envelope progenitor stars that experience undergo late-time interaction with CSM shells, as well as compare its evolutionary state to the young O-rich SNR NGC 4449-1, for which we have taken new data. Our optical and X-ray spectral data and results are described in $\S$2 and $\S$3, with discussion and conclusions given in $\S$4 and $\S$5.


\section{Observations}

\subsection{Optical Spectra}

Low-resolution optical spectra of SN~1996cr were obtained on three nights: 2017 July 19, and 2021 August 1 and August 2 using the 6.5m Baade Magellan Telescope at Las Campanas Observatory with the Inamori Magellan Areal Camera and Spectrograph (IMACS). This spectrograph employs a mosaic of eight 2k $\times$ 4k CCDs in an array mounted behind an f/4 camera providing a 15.4 arcmin FOV.  Two 2400 second exposures were obtained on each night using a 0.9 arcsecond wide N-S slit with a 300 lines mm$^{-1}$ grating (0.746 \AA \ pixel$^{-1}$) centered at 7300 \AA. 

Wavelength coverage was 4200 -- 10400 \AA. Seeing was excellent in 2017 with a full width half maximum (FWHM) = $0\farcs5$ to $0\farcs6$ but with some light cirrus, while seeing in 2021 ranged from $0\farcs7$ to $1\farcs1$.  The spectra from each night were co-added to form a single 4800 second exposure. Spectral resolution (FWHM) as measured from comparison lamp spectra was 5.85 and 5.60 \AA \ at 5000 and 6700 \AA, respectively.  Both data sets were redshifted corrected by +560 km s$^{-1}$ of that of the faint coincident H~II region. 

While the 2017 and 2021 exposure times were identical, the 2021 flux level was about half that seen in 2017. Thus the S/N of the 2017 data are better than the 2021 data. Because of this and the small changes seen between the 2017 and 2021 spectra, in the analysis that follows below we will mainly focus on the 2017 spectrum. 

Spectra were also taken on 2021 July 12 using the Robert Stobie Spectrograph (RSS) on the 10~m SALT telescope in South Africa. Using a 900 lines per mm grating and a $1\farcs5$ wide slit, spectra were obtained covering 4050 -- 7400 \AA \ region with a  FWHM resolution of 5 \AA \  and a dispersion of 1.0 \AA\ pix$^{-1}$. Two 1200 s exposures were obtained under $1\farcs5$ to $2\farcs0$ seeing conditions. Although these data were subsequently superseded by the follow-up Magellan observations they were used to confirm the major features of the 2021 spectrum of SN~1996cr.

We also examined archival, moderate-resolution late-time spectra of SN~1996cr obtained at the VLT using the X-Shooter spectrograph \citep{Vernet2011}. These data were taken on 05 May 2016 and 01 May 2019  with exposures of $2 \times 1052$ s and $2 \times 1500$ s, respectively and covered the instrument's UBV/VIS wavelength regions (3000 - 10 200 \AA). Due to the higher resolution of these data (R $\simeq$ 8900), we have used these to better examine the width and profile of SN~1996cr's broad H$\alpha$ emission.

As mentioned previously, we also obtained low-dispersion spectra of the young $\lesssim$ 100 yr O-rich SNR in NGC~4449 for the purposes of comparison with SN~1996cr \citep{Blair1983,Mili2008,Bietenholz2010}. These spectra were obtained in May 2019 and February 2020 with the 2.4m Hiltner telescope at MDM Observatory using the Ohio State Multi-Object Spectrograph (OSMOS: \citealt{Martini2011}). Spectra taken at both epochs were obtained using 1.4\arcsec\ wide N-S slit and a red VPH grism (R = 1600 at 8000 \AA) with a wavelength coverage of $4000 - 8600$ \AA. For the 2020 spectra, a second set of spectra where taken to cover $5000 -10000$ \AA.  Both spectral setups yielded a measured spectral resolution of 7 \AA. Two sets of 2000 s exposures where taken in both set-ups at both epochs which were then co-added into separate 4000 s exposures.

Standard pipeline data reduction of  Magellan and MDM spectra using AstroPy and PYRAF\footnote{PYRAF is a product of the Space Telescope Science Institute, which is operated by AURA for NASA.} were performed.  Spectra were reduced using the software L.A.\ Cosmic \citep{vanDokkum01} to remove cosmic rays.  Spectra were calibrated using Ar, He, and Ne lamps and spectroscopic standard stars \citep{Oke74,Massey90,Hamuy1992,Hamuy1994}. SALT spectra were reduced using specific SALT reduction software. 

\subsection{X-Ray Observations}

Following its serendipitous discovery \citep{Bauer2001} (see the X-ray image shown in Fig.~\ref{fig:tcolor}) SN~1996cr was observed extensively with {\it Chandra}, {\it XMM--Newton}, and {\it Swift} \citep{Bauer2008}. As seen in Figure~\ref{fig:tcolor}, SN~1996cr lies approximately 20$\arcsec$ from the X-ray bright active nucleus of the Circinus Galaxy. Its location relative to other X-ray sources in the galaxy allows for detailed X-ray spectral studies with \textit{Chandra}, with little contamination due to nearby X-ray bright sources. Given the extraordinary evolution of its X-ray and optical emission \citep{Bauer2008,Dewey2011}, we were granted \textit{Chandra} Guaranteed Time from the High Resolution Camera Instrument Team with observations performed in late November 2018. 

SN~1996cr was observed for 50 ksec (Table~\ref{tab:XRAY}) with ACIS-S in very faint (VF) timed exposure mode. SN~1996cr was also observed with ACIS-S previously in 2000, 2004, 2008, and 2010. The data from each observation was reprocessed using \texttt{CIAO} version $4.13$ and version $4.9.5$ of the \textit{Chandra} CALDB. Spectra and associated response files were extracted or generated using the \texttt{CIAO} tool \texttt{spec\_extract}. For those observations from 2008 which employed the \textit{Chandra} High Energy Transmission Grating (HETG), we extracted only the zeroth order spectrum. A detailed analysis of the dispersed spectrum is discussed in \cite{QV2019}. The data were binned to a minimum of 15 counts per PHA channel, and fit using \textsc{Xspec} version 12.10.1f. A description of the spectral fits and results can be found in Section~\ref{sec:cxcxray}.

\begin{figure*}
\centering
\includegraphics[width=0.9\textwidth]{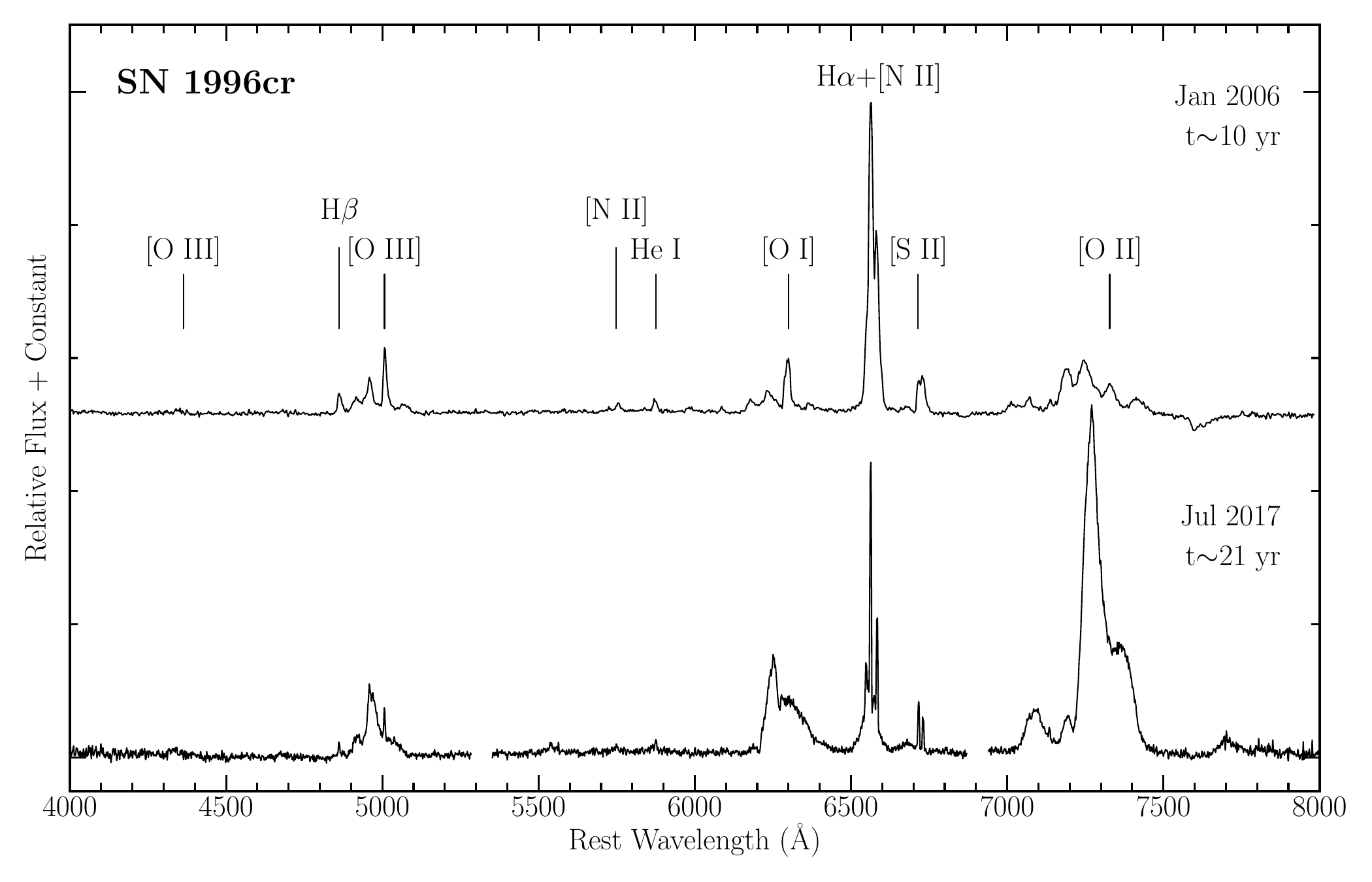}
\caption{A comparison of a January 2006 VLT spectrum of SN~1996cr at an age $\sim$ 10 years from \citet{Bauer2008} compared with our $\sim$21 year Magellan spectra showing the radical change in the spectrum of SN~1996cr between ages 10 and 21 years ($\sim$ first and decades post-max). Line identifications for the strongest emission features in the 2006 spectrum are marked. The gaps in the Magellan spectrum from 2017 at $\sim$ 5300 \AA \ and 6900 \AA \ are due to chip gaps in the IMACS detector.}
\label{Bauer_comp}
\end{figure*}

%


\section{Results}

\begin{deluxetable*}{llllcccccccc}[ht]
\tablecolumns{12}
\tablecaption{Deblended SN1996cr Line Emissions at t $\simeq$ 10 versus 21.5 Years 
\label{Table_deblend}}
\tablewidth{0pt}
\tablehead{\colhead{} & \colhead{} & \colhead{} & \multicolumn{4}{c}{t $\simeq$ 10 years\tablenotemark{a}}  & \colhead{}   &  \multicolumn{4}{c}{t $\simeq$ 21.5 years} \\
\cline{4-7} \cline{9-12}
\colhead{Emission} & \colhead{Line} & \colhead{$\lambda_{\rm lab}$} & \colhead{$\lambda_{\rm obs}$} & \colhead{${\rm V_{peak}}$} &\colhead{FWHM\tablenotemark{b}}  & \colhead{Inferred} & \colhead{} 
& \colhead{$\lambda_{\rm obs}$}  & \colhead{${\rm V_{peak}}$}  & \colhead{FWHM\tablenotemark{c}} & \colhead{Inferred} \\
\colhead{Component}  & \colhead{ID} & \colhead{(\AA)}  & \colhead{(\AA)} &  \colhead{(km s$^{-1}$)} & \colhead{(km s$^{-1}$)} & \colhead{Source} & \colhead{}  
& \colhead{(\AA)} & \colhead{(km s$^{-1}$)} & \colhead{(km s$^{-1}$)}  &  \colhead{Source}   }
\startdata
{\bf Narrow}\tablenotemark{d} & H$\alpha$ & 6563 & $6563.0\pm0.2$ & $\simeq 0$ & $669\pm18$ & SN-CSM 
& ~~ & $6562.8\pm0.2$ & $0 \pm 20$ & $50 \pm 15 $ &  H II Region \\
 ~~~~ " & H$\beta$     & 4861 & $4861.1\pm0.2$ & " &  "~~ "      & "~" & & $4861.5\pm0.4$ &  "  & $< $ 60   &  " ~ ~ " ~ ~ "  \\
 ~~~~ " & He I         & 5876    & $5873.9\pm0.8$ & " &  $913\pm65$ & "~" & & $5875.4\pm0.4$ &  "  & \nodata  &  " ~ ~ " ~ ~ "  \\
 ~~~~ " & $[$S~II$]$   & 6731 & $6732.6\pm0.8$ & " & $741\pm24$  & "~" & & $6730.8\pm0.2$ &  "  & $50 \pm 15$  &  " ~ ~ " ~ ~ "  \\
 ~~~~ " & $[$S~II$]$   & 6716 & $6717.6\pm0.8$ & " &  "~~ "      & "~" & & $6716.5\pm0.2$ &  "  &  " &  " ~ ~ " ~ ~ "  \\
 ~~~~ " & $[$N~II$]$   & 6548 & $6548.8\pm0.2$ & " &  "~~ "      & "~" & & $6547.8\pm0.3$ &   " &  "  &  " ~ ~ " ~ ~ "  \\
 ~~~~ " & $[$N~II$]$   & 6583 & $6582.8\pm0.2$ & " &  "~~ "      & "~" & & $6583.4\pm0.3$ &   " & "   &  " ~ ~ " ~ ~ "  \\
 ~~~~ " & $[$O~III$]$  & 5007 & $5007.9\pm0.3$ & " &  "~~ "      & "~" & & $5006.6\pm0.3$ &   " & $< $ 60 &  " ~ ~ " ~ ~ "  \\ 
 ~~~~ " & $[$Ar~III$]$ & 7136 & $7136.8$~~~    & " &  "~~ "      & "~" & & $7135.4\pm0.5$ &   " & \nodata &  " ~ ~ " ~ ~ "  \\
{\bf Broad}\tablenotemark{e} & H$\alpha$    & 6563 &$6563.0\pm0.2$ & $\sim 0$ & $4077\pm398$ & SN ejecta &  &  6558 & ~ $-220$ &$2600\pm100$ & CSM \\
red/c1  & $[$O I$]$    & 6300    & 6376.9         & +3620   &  $2049\pm19$  &  SN ejecta & &  6326   & $+1220$  & $4100\pm100$ & SN ejecta  \\
red/c1  & $[$O II$]$   & 7325    & 7420.1         & +3890   &  $2049\pm19$  &  " ~~ "  & &  7355  & $+1220$  & $4100\pm100$ &  " ~~ " \\ 
red/c1  & $[$O III$]$  & 5007    & $5066.4\pm1.1$ & +3360   &  $2049\pm19$  &  " ~~ "  & &  5027 & $+1220$  & $4100\pm100$ &  " ~~ " \\
red/c1  & $[$Ar III$]$ & 7136    & \nodata        & \nodata &  \nodata      &  \nodata & &  7189  & $+2200$  & $1300\pm100$  & " ~~ " \\
red/c1  & $[$S III$]$  & 9531    & \nodata        & \nodata &  \nodata      &  \nodata & &   9596 & $+2050$  & $2900\pm100$ &  " ~~ " \\
blue/c2 & $[$O I$]$    & 6300    & 6231.6         &$-3260$  &  $2049\pm19$  &  " ~~ "  & &  6251 & $-2330$  & $2300\pm100$  & " ~~ " \\
blue/c2 & $[$O II$]$   & 7325    & 7251.1         &$-3030$  &  $2049\pm19$  &  " ~~ "  & &  7268  & $-2330$  & $2300\pm100$  &  " ~~ " \\
blue/c2 & $[$O III$]$  & 5007    & $4952.6\pm0.6$ &$-3260$  &  $2049\pm19$  &  " ~~ "  & &  4968 & $-2330$  & $2300\pm100$  &  " ~~ " \\
blue/c2 & $[$Ar III$]$ & 7135    & \nodata        & \nodata &  \nodata      &  \nodata & &  7090 & $-2000$  & $3000\pm100$ &  " ~~ " \\
blue/c2 & $[$S II$]$   & 6716    & \nodata        & \nodata &  \nodata      &  \nodata & &  6679 & $-2000$  & $2300\pm100$ &  " ~~ " \\
blue/c2 & $[$S III$]$  & 9531    & \nodata        & \nodata &  \nodata      &  \nodata & &  9468 & $-2000$  & $2500\pm100$ &  " ~~ " \\
blue/c3 & $[$O I$]$    & 6300    & 6179.6         & $-5730$ &  $2049\pm19$  &  " ~~ "  & & \nodata  & \nodata  & \nodata & \nodata  \\
blue/c3 & $[$O II$]$   & 7325    & 7190.5         & $-5500$ &  $2049\pm19$  &  " ~~ "  & & \nodata  & \nodata  & \nodata & \nodata  \\
blue/c3 & $[$O III$]$  & 5007    & $4910.8\pm0.6$ &$-5760$  &  $2049\pm19$  &  " ~~ "  & & \nodata  & \nodata  & \nodata & \nodata  \\ 
\enddata 
\tablenotetext{a}{Data for t $\simeq$ 10 years taken from \citet{Bauer2008}.} 
\tablenotetext{b}{\citet{Bauer2008} narrow component FWHM velocities uncorrected for instrumental resolution.}    
\tablenotetext{c}{Our narrow component FWHM velocities corrected for instrumental resolution. }
\tablenotetext{d}{Comparison of narrow velocity components are restricted to the zero velocity components that were also found in \citet{Bauer2008}.}
\tablenotetext{e}{Redshifted or blueshifted components indicated by ``red" or ``blue", along with \citet{Bauer2008} component identifications: ``c1", ``c2", and ``c3". Velocity values for our 21.5 yr data are from deblending model fits. }
\end{deluxetable*}


Our July 2017 and August 2021 spectra of SN~1996cr when it was $\simeq$ 21 and 25 yr old look radically different from that observed in early 2006 by \cite{Bauer2008} when SN~1996cr was around 10 yr old. 
Figure~\ref{Bauer_comp} shows a
comparison of SN~1996cr's 2006 spectrum taken by \citet{Bauer2008} to our 2017 spectrum.
For this comparison, a strong and highly reddened continuum has been removed from the  \citet{Bauer2008} spectrum. Moreover, while our 2017 spectrum covered 4000 to  10~000 \AA, for this comparison we show only the wavelength region of $4000 - 8000$ \AA \ to match the data of \citet{Bauer2008}.  In viewing this comparison  it is important to note that the \citet{Bauer2008} 2006 spectrum is of lower spectral resolution compared to our 2017 spectrum ($12 - 13$ \AA \  vs.\ $5 - 6$ \AA), thereby significantly affecting a simple comparison of emission line widths between the two epochs. 

Nonetheless, Figure~\ref{Bauer_comp} highlights the significant and surprising changes in SN~1996cr's late-time optical properties at ages $\simeq$ 10 and 21 years after outburst. Most notable are the changes in the widths of the profiles of the forbidden oxygen emission lines, along with changes in the brightness of these emissions relative to those of H$\alpha$ and [\ion{N}{2}]. Moreover, whereas in 2006 SN~1996cr showed multiple blue and red shifted components, a decade or so later only single blue and red components appear present. We discuss these changes in the velocity profiles of the lines in more detail in Section 3.1. Overall, the spectrum of SN~1996c spectrum at 21 yr old exhibits much broader emission components compared to its spectrum at 10 yr old but with much lower absolute peak velocities.

\begin{figure*}
\centering
\includegraphics[width=0.95\textwidth]{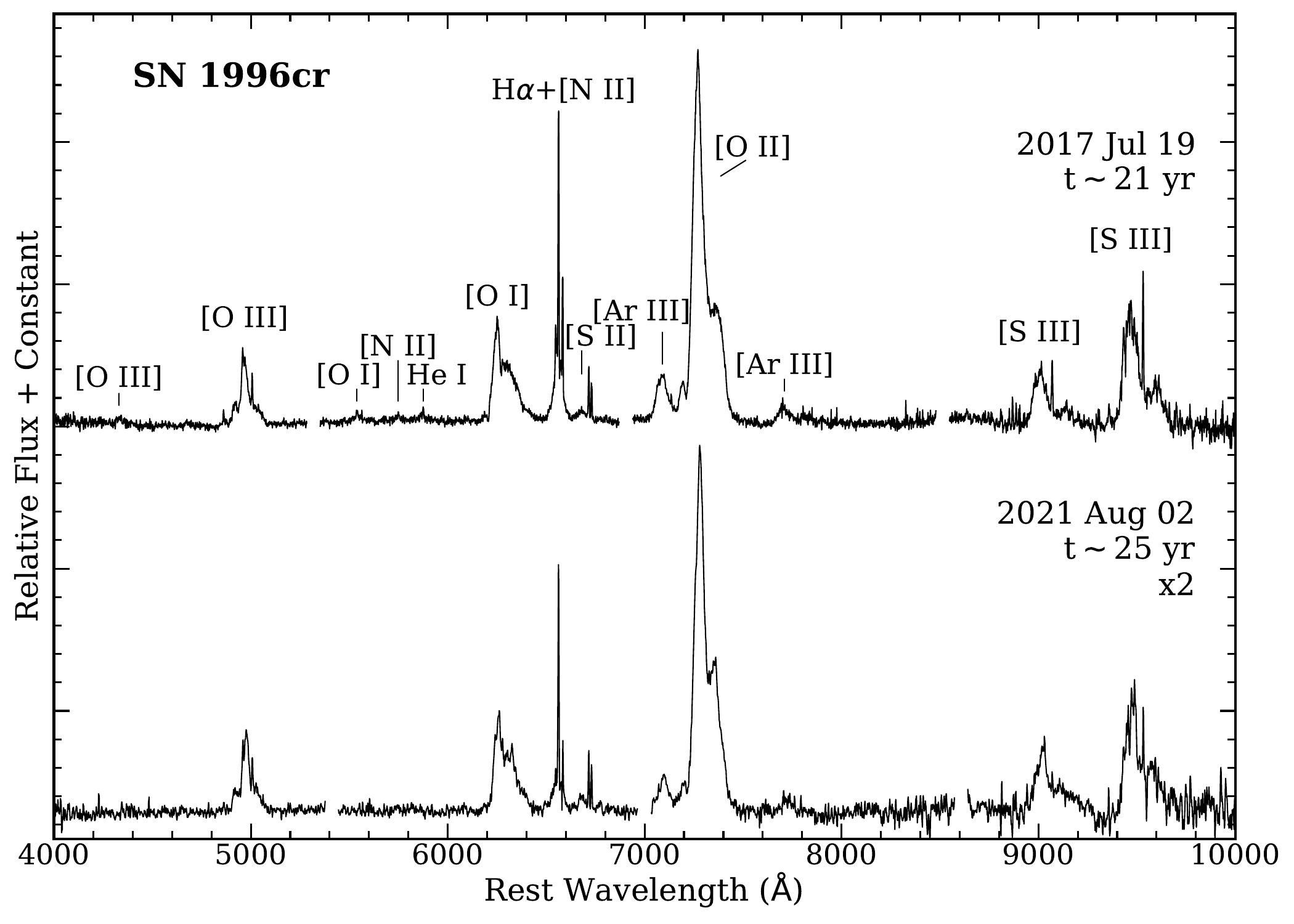}
\caption{Magellan IMACS spectra of SN~1996cr obtained approximately 21 and 25 years after estimated date of outburst. }
\label{opt_spectrum}
\end{figure*}

\citet{Bauer2008} deblended their 2006 spectrum into two or three emission components assuming Gaussian profiles. We have chosen to follow a similar procedure for deblending our spectra into different emitting components. We chose to decompose the spectrum into contributions from shocked ejecta, which is represented by broad blue and red-shifted emission, and narrow components which we assume are related to emission from shocked circumstellar material. For the oxygen emission lines, we choose to use the [\ion{O}{2}]  feature as it was visually easier to disentangle the components compared to [\ion{O}{3}]. Our model fits for the 21 year spectrum differ from those done on the 10 yr spectrum, as the red and blue components were not required to have the same FWHM and the 10 yr fits were based on [\ion{O}{3}]. The results of observed and model fits to our 2017 spectrum are listed in Table~\ref{Table_deblend}, along with values taken from \citet{Bauer2008}. In this table, we have separated narrow from broad emission features, with the broad emission further divided into red and blue shifted components following the naming convention of \citet{Bauer2008}.

As expected from the comparison of the 2006 and 2017 spectra of SN~1996cr shown in Figure~\ref{Bauer_comp}, Table~\ref{Table_deblend} reveals large differences between 10 and 21 yr FWHM values for SN~1996cr's especially for the narrow emission lines. For the 2006 spectrum, \citet{Bauer2008} lists values of 669 km s$^{-1}$ for H$\alpha$, 913 km s$^{-1}$ for \ion{He}{1} with a value of 741  km s$^{-1}$ for all other narrow line emissions. In contrast, we find FWHM values for the 2017 spectrum all narrow lines with FWHMs below 60 km s$^{-1}$.

Our 2017 spectrum showed narrow line widths only slightly above instrumental resolution, whereas \citet{Bauer2008} quotes values seemingly uncorrected for their much lower instrumental spectral resolution. While \citet{Bauer2008} suggest that the narrow feature may be marginally resolved above the instrumental uncertainty, this leads to uncertainty concerning if there has been significant line width changes in the relatively narrow emission features seen in the optical spectrum over the last decades. The low resolution of the 2006 spectrum also leaves unanswered whether there was narrow blue shifted  H$\alpha$ and [\ion{N}{2}] line emissions at age 10 years like that seen in the 21 yr spectrum.

Table~\ref{Table_deblend} also highlights the large differences for the broad line emissions during SN~1996cr's late-time evolution between ages 10 and 21 years. For example, the very high blue shifted emissions, labeled component `blue/c3' seen in 2006 is entirely absent in the 2017 spectrum. In addition, another blue shifted component, `blue/c2', has a lower emission peak velocity while being noticeably broader in velocity. Also, as already noted above, the emission peak velocities of the oxygen lines are $\simeq$300 km s$^{-1}$ higher than for sulfur and argon, we could not discern a difference in FWHM values for this component.

The sole redshifted component, `red/c1', likewise underwent a significant evolution. \citet{Bauer2008} found a peak velocity around 3600  km s$^{-1}$ for the oxygen emissions in 2006, but by 2017 this dropped to just over 1200 km s$^{-1}$. On the other hand, SN~1996cr's red shifted oxygen emission profiles became much broader with a FWHM increasing from $\sim$2000 to 4100 km s$^{-1}$.

In our 2017 spectrum, redshifted peak sulfur and argon emissions exhibit higher velocities than oxygen emissions, just the opposite of that seen for the broad blueshifted lines. On the other hand, these sulfur and argon lines show much narrower line widths than the oxygen lines; 1300 -- 2900 km s$^{-1}$ vs.\ 4100 km s$^{-1}$.

Lastly, SN~1996cr's hydrogen emission also changed significantly between ages 10 and 21 years. \citet{Bauer2008} found broad H$\alpha$ emission with a FWHM $\simeq4100$ km s$^{-1}$ centered around zero rest velocity. In contrast, our 2017 spectrum showed a much lower, intermediate-width FWHM around 2600 km s$^{-1}$ and slightly blueshifted at $-220$ km s$^{-1}$. Such a difference in the H$\alpha$ line width is not clear from an examination of the  \citet{Bauer2008} data (see Fig.~\ref{Bauer_comp}) and we are unable to confirm the presence of the reported $\simeq4100$ km s$^{-1}$ component, in the archival data that we inspected. However, what is clear is that intermediate-width hydrogen emission is present in the 2017 late-time spectrum of SN~1996cr, presumably due to shocked and accelerated pre-SN mass loss material. 


\subsection{SN~1996cr's Optical Spectral Evolution from 2017--2021}

\begin{figure*}
\centering
\includegraphics[width=0.42\textwidth]{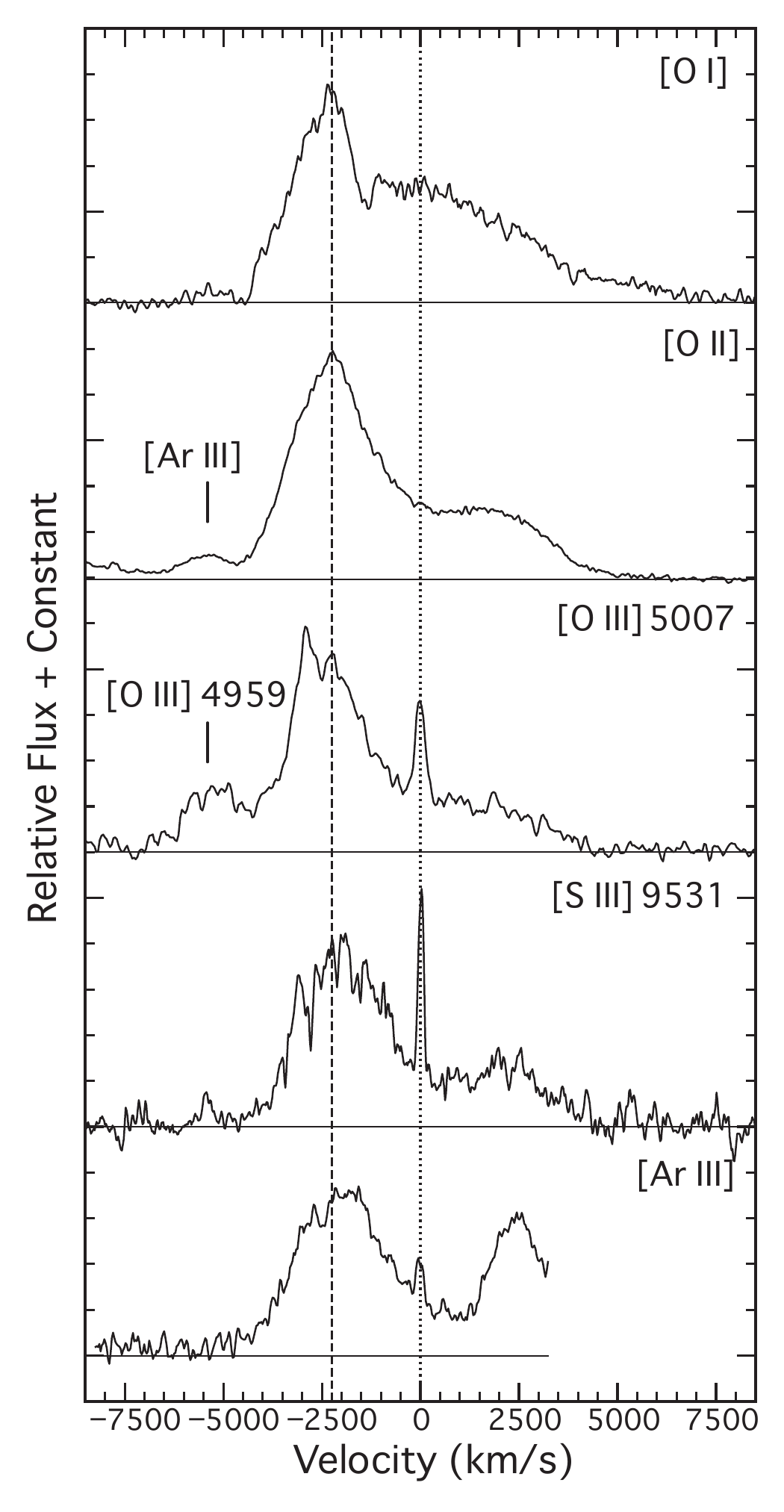}
\includegraphics[width=0.42\textwidth]{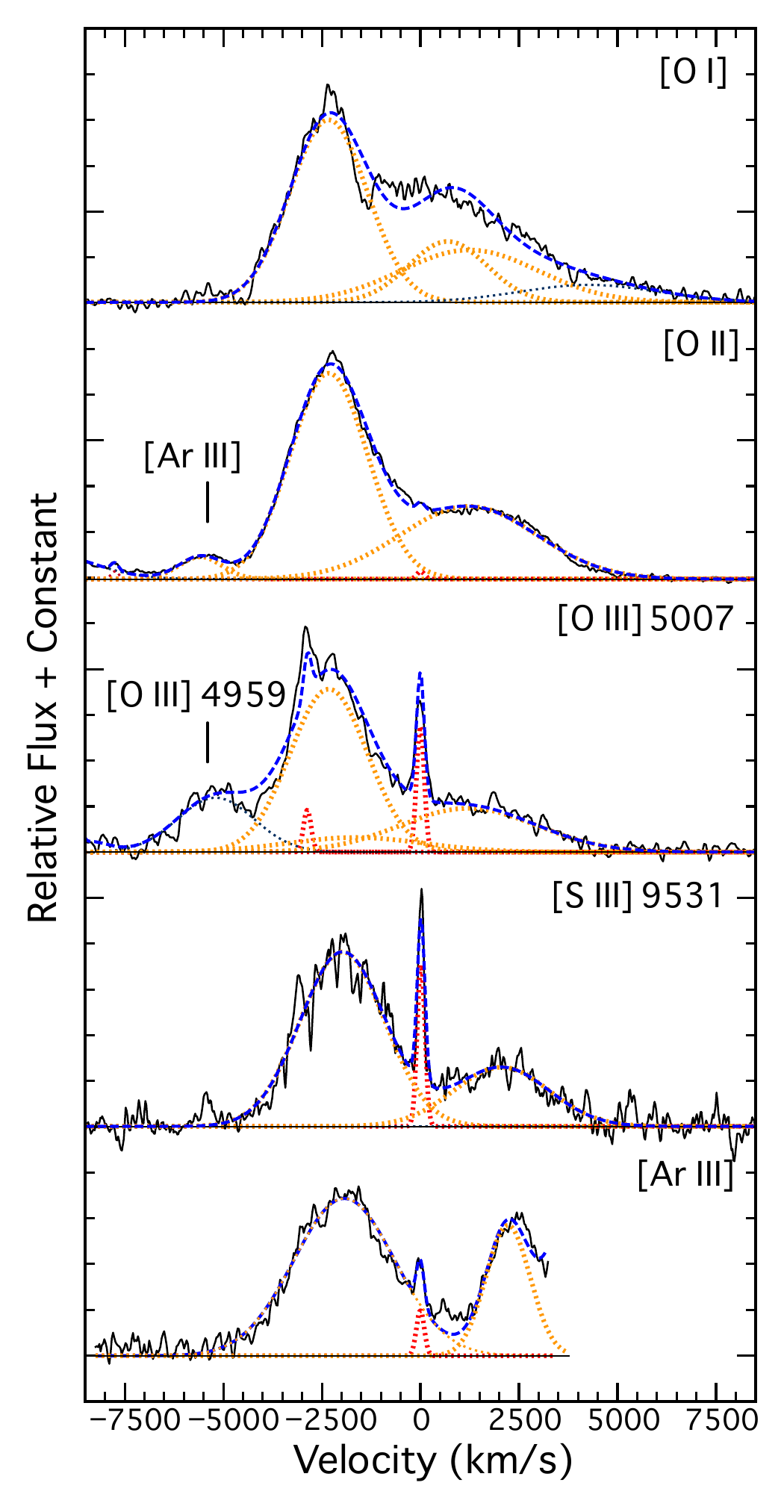}
\caption{Plots of broad line emissions relative to expansion velocity observed in July 2017 for [\ion{O}{1}] $\lambda$6300, [\ion{O}{2}] $\lambda$7325 [\ion{O}{3}] $\lambda$5007, [\ion{S}{3}] $\lambda$9531, and [\ion{Ar}{3}] $\lambda$ line emissions. Left-hand plots: Vertical black dotted lines mark the locations of zero rest frame while the black dashed line marks the $-2250$ km s$^{-1}$ velocity peak for the oxygen emission lines. Right-hand plots: Deblending model fits to the emission line profiles are shown using Gaussian components. The full model is shown as a blue dashed line. Both broad red and blue shifted ejecta emissions shown as dotted orange lines, while the narrow components from the surrounding H~II region are shown as dotted red lines. 
\label{O_vel}}
\end{figure*}

A comparison of SN~1996cr's optical spectra in July 2017 and August 2021 is shown in Figure~\ref{opt_spectrum} where major emission features  are identified.  Adopting an explosion date in early 1996, these spectra show SN 1996cr's optical emissions at an approximate age of 21.5 and 25.5 years. Gaps in the spectra near 5300 - 5400 \AA, 6900 - 7000 \AA, and 8560 - 8650 \AA \ are due to spaces in the IMACS spectrograph's mosaic of CCD chips.  

Broad emission features are seen due to forbidden oxygen, sulfur, and argon emissions. These include [\ion{O}{1}]$\lambda\lambda$6300,6364, [\ion{O}{2}] $\lambda\lambda$7319,7330, [\ion{O}{3}] $\lambda\lambda$4959,5007, [\ion{S}{2}] $\lambda\lambda$6716,6731, [\ion{S}{3}] $\lambda\lambda$9069,9531, and [\ion{Ar}{3}] $\lambda$7136, along with less broad H$\alpha$ emission. Narrow emission lines are also detected presumably associated with the diffuse H~II region seen coincident with and immediately adjacent to SN~1996cr (see Fig.\ 1). These include H$\alpha$, [\ion{O}{3}], \ion{He}{1} $\lambda$5876,  [\ion{N}{2}] $\lambda\lambda$6548,6583, and the [\ion{S}{2}], and [\ion{S}{3}] doublets.

With the exception of H$\alpha$, two distinct blueshifted and redshifted emission components are seen for all broad line emissions. The signature of internal dust extinction in SN~1996cr's ejecta is likely the source of the noticeable strength difference between blue- and redshifted emission components, most obvious in the [\ion{O}{2}] $\lambda\lambda$7319,7330 and [\ion{S}{3}] $\lambda\lambda$9069,9531 line emissions.

\begin{figure*}
\centering
\includegraphics[width=0.42\textwidth]{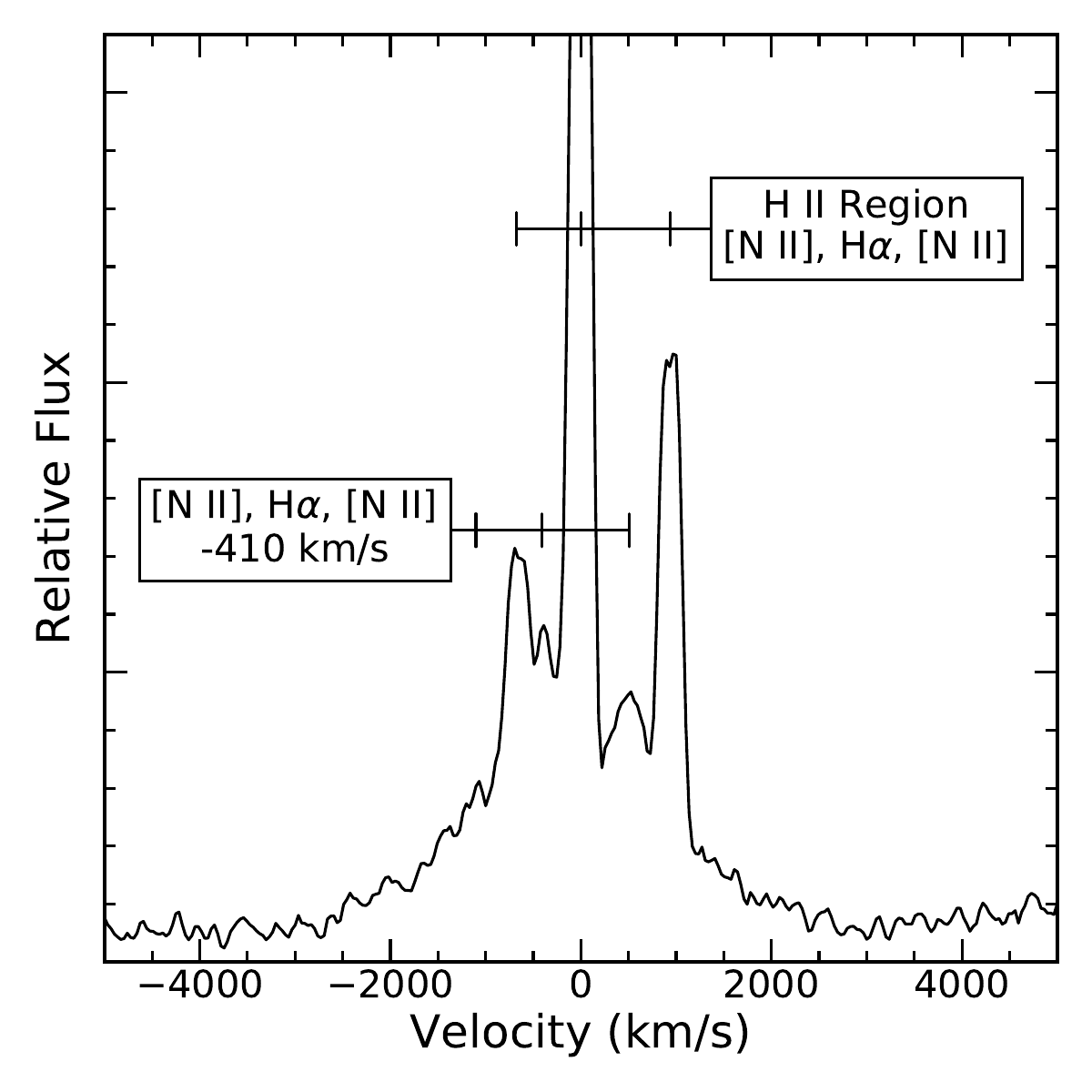}
\includegraphics[width=0.42\textwidth]{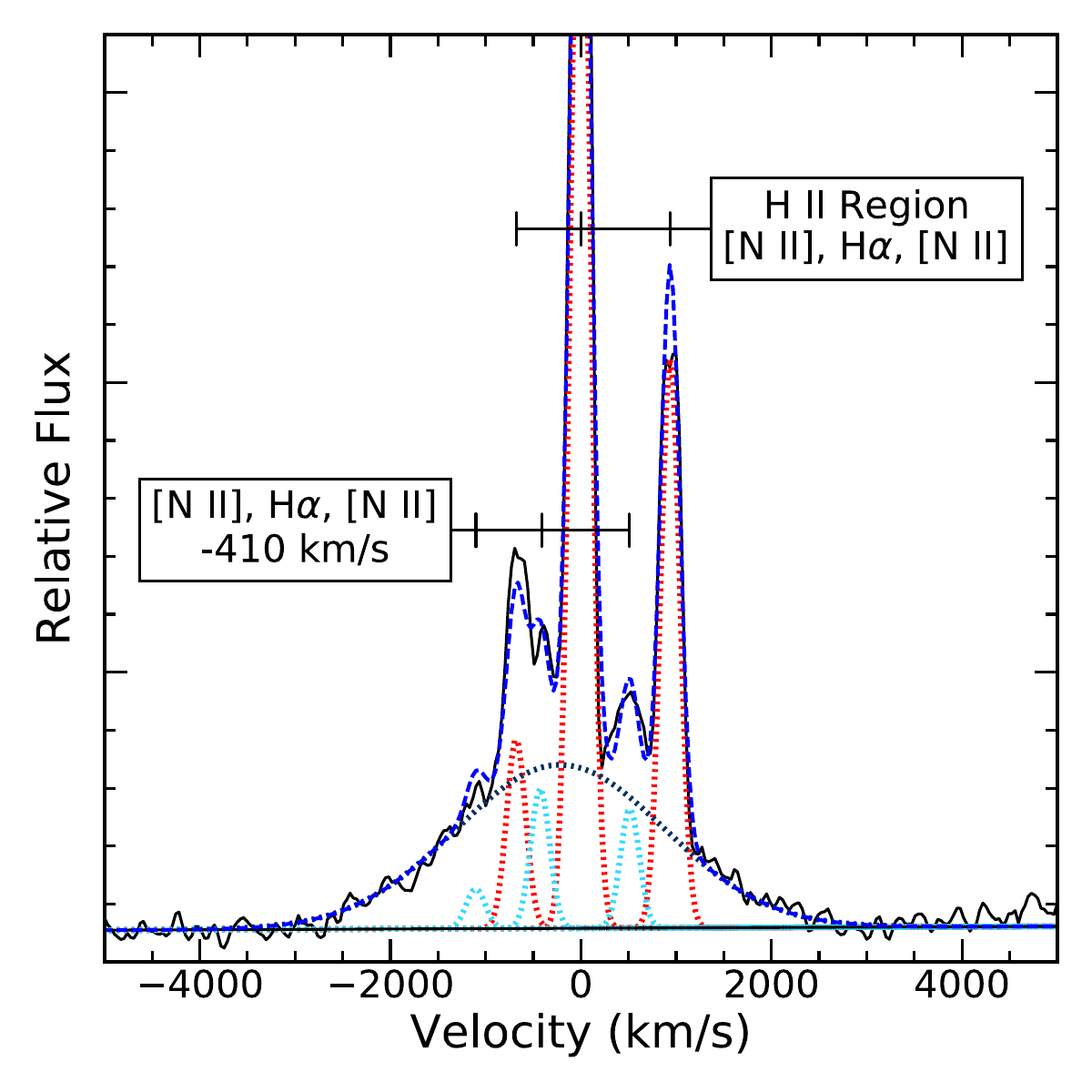}
\caption{Decomposition of emission centered around H$\alpha$ observed in our July 2017 spectrum of SN\,1996cr. Left: Observed emission with blended contributions from H$\alpha$ and [\ion{N}{2}] $\lambda\lambda$6548, 6583 lines associated with SN\,1996cr and the \ion{H}{2} region.  
Right: Deblending model fits to the emission line profiles using Gaussian components. The total model is shown as a dotted blue line. The emission from the \ion{H}{2} region is shown as a dotted red line, while the emission from the $-410$ km s$^{-1}$ component is shown as a dotted cyan line. The broad H$\alpha$ emission (shown as a black dotted line), centered at $-220$~km\,s$^{-1}$ and spanning full-width-at-zero-intensity velocities of $-2800$ to $+2600$~km\,s$^{-1}$, is associated with SN-CSM interaction.}
\label{fig:H_vel}
\end{figure*}

Enlargements of the broad emission features seen in the 2017 spectrum are presented in the left-hand panels of Figure~\ref{O_vel} were we show a series of radial velocity plots of the observed broad emissions of oxygen, sulfur and argon. A dashed vertical line in the left hand panels marks the emission peak velocity at $-2250 \pm 75$ km s$^{-1}$ for the three oxygen line emissions. Broad sulfur and argon emissions exhibit noticeably lower peak velocities than that of the oxygen lines, roughly 300 km s$^{-1}$ less, peaking at around $-1950$ km s$^{-1}$.  Narrow line emissions from the coincident H~II region are visible in these plots for [\ion{O}{3}] $\lambda\lambda$4959, 5007, [\ion{S}{3}] $\lambda$9531, and [\ion{Ar}{3}] $\lambda$7136 and these are marked with a dotted line at zero velocity equal to the rest frame velocity of local H~II region in the Circinus galaxy.

\begin{deluxetable*}{llccccccccc}
\tablecolumns{10}
\tablewidth{0pc}
\tablecaption{Observed SN 1996cr July 2017 Emission Line Velocities, Fluxes and Luminosities 
\label{table_1}}
\tablehead{
\colhead{Line} & \colhead{$\lambda_{\rm lab}$} & \colhead{Emission} & \colhead{$\lambda_{\rm peak}$} &  \colhead{V$_{\rm peak}$} & 
                     \colhead{$\lambda_{\rm FWZI}$}    &  \colhead{V$_{\rm FWZI}$} & 
                     \colhead{Observed Flux} & \colhead{Corrected Flux\tablenotemark{a}} & \colhead{Luminosity\tablenotemark{b}}  \\
\colhead{ID}     & \colhead{(\AA)}  & \colhead{Peak}   & \colhead{(\AA)}               & \colhead{(km s$^{-1}$)} & 
                     \colhead{(\AA)} & \colhead{(km s$^{-1}$)} &
        \colhead{(10$^{-15}$ erg s$^{-1}$)} & \colhead{(10$^{-14}$ erg s$^{-1}$)} &  \colhead{(10$^{37}$ erg s$^{-1}$)} 
}
\startdata
  H$\alpha$    & 6563         & \nodata   & 6558   & ~ $-220 \pm 75$ & 6502 - 6620 & $-2800 ~ +2600$  & 0.8 & 0.7 &  1.5 \\
  $[$O I$]$    & 6300,6364    &  blue     & 6250   & $-2380 \pm 50$  & 6207 ~~ \nodata & $-4430$ ~~~ \nodata  & 6.4   &  5.9 & 12.4 \\
  $[$O III$]$  & 4363         &  blue     & 4330   & $-2250 \pm 90$  &   \nodata   & \nodata          & 0.1  &  0.3 &  0.6 \\
  $[$O II$]$   & 7319,7330    & blue      & 7270   & $-2250 \pm 75$  & 7215 ~~ \nodata & $-4500$ ~~~ \nodata  &16.8~~ & 10.3 & 21.7 \\
  $[$O III$]$  & 4959,5007    &  blue     & 4970   & $-2270 \pm 75$  & 4890 ~~ \nodata & $-4200$ ~~~ \nodata & 3.0  &  6.2 & 13.0 \\
  $[$S II$]$   & 6716,6731    &  blue     & 6679 & $-1960 \pm 75$    & 6627 ~~ \nodata & $-4000$ ~~  \nodata & 0.2  &  0.2 &  0.4 \\
  $[$Ar III$]$ & 7136         & blue      & 7090 & $-1950 \pm 75$    & 7035 - 7160    & $-4250$  +1000  &  2.1  &  1.5 &  3.2 \\
  $[$Ar III$]$ & 7751         & blue      & 7702 & $-1900 \pm 75$    & 7650 - 7770    & $-3910$ ~  +735 &  0.8  &  0.4 & 0.8  \\
  $[$S III$]$  & 9069         & blue      & 9008 & $-2000 \pm 50$    & 8940 ~~ \nodata  & $-4270$ ~~ \nodata  &   2.6 &  1.0 &  2.1 \\
  $[$S III$]$  & 9531         & blue      & 9469 & $-1950 \pm 50$    & 9395 ~~ \nodata & $-4280$ ~~ \nodata &   7.7  & 2.4 &  5.0 \\
  \hline
  $[$O II$]$   & 7319,7330    & red       & 7360 & $+1450 \pm 50$    & \nodata ~~ 7445  & \nodata  ~~ +4700  & 6.3  & 3.9 & 8.1 \\
  $[$S III$]$  & 9069         & red       & 9135 & $+2190 \pm 75$    & \nodata ~~ 9175 & \nodata ~~ +3500 &  0.6   & 0.2 &  0.4 \\
  $[$S III$]$  & 9531         & red       & 9600 & $+2170 \pm 75$    & \nodata ~~ 9670 & \nodata ~~ +4400 & 2.2    & 0.6 &  1.3 \\
  $[$Ar III$]$ & 7136         & red       & 7193 & $+2400 \pm 75$    & 7165 ~~ \nodata  & $-1450$ ~~ \nodata &  0.7   & 0.6 &  1.3 \\ 
\enddata
\tablenotetext{a}{Observed fluxes corrected assuming an $E(B-V)$ = 0.91.}
\tablenotetext{b}{Values assume a distance of 4.2 Mpc.}
\end{deluxetable*} 

\subsubsection{Emission-Line Velocities}

In Table~\ref{table_1}, we list SN~1996cr's July 2017 observed peak emission line wavelengths and radial velocities, emission width at zero intensity (FWZI) wavelengths and radial velocities, along with observed and extinction corrected line fluxes and luminosities.  Broad emissions of oxygen, sulfur, and argon  measurements are subdivided into features showing distinct blue and red emission peaks.  For both groups, derived peak radial velocity values assumed the line lab wavelength shown and the observed peak wavelengths listed.

Peak wavelengths were measured using IRAF profile routines, except for the broad H$\alpha$ feature which was determined by Gaussian line profile fitting. For the listed FWZI velocities for emission doublets, we took the bluer or redder line component in calculating the maximum blue or red velocities. Observed flux values are believed accurate only to $\pm$30\% due to slit losses and variable atmospheric conditions.


Blending of blue and red components prevented measurements of some FWZI values and corresponding radial velocities.  This is reflected in Table~\ref{table_1} where only the short wavelength side of FWZI values are listed for the blueshifted emission components with the exception of the [Ar III] $\lambda$7136 line. Similarly, only the red side of the FWZI values for the redshifted components were measurable. 
 
These line measurements reveal important velocity differences between oxygen velocities and those of sulfur and argon. Blueshifted oxygen emission peak velocities all lie at around $-2250$ km s$^{-1}$ which is roughly 300 km s$^{-1}$ faster than the $-1950$ km s$^{-1}$ velocity measured for blue-shifted sulfur and argon emission peaks. Interestingly, measured redshifted emission peak velocities for [\ion{O}{2}] are lower than those for sulfur and argon. This may provide some clues as to the asymmetry in the explosion.

\begin{figure}
    \includegraphics[width=0.48\textwidth]{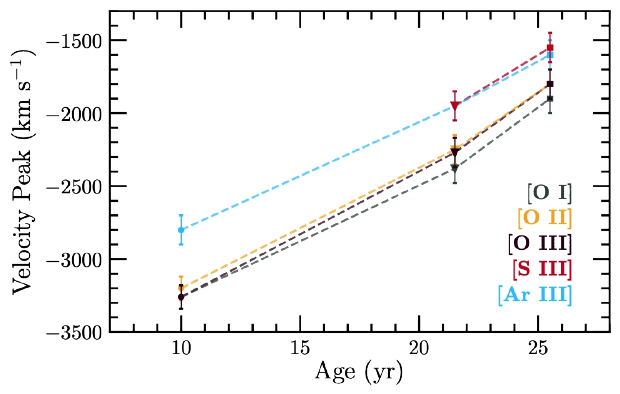}
    \caption{Evolution of SN~1996cr's O, S, and Ar emission peak velocities 10, 21.5, and 25.5 years after estimated date of  outburst.}
    \label{fig:oxy_evolution}
\end{figure}

This difference is shown in Figure~\ref{fig:oxy_evolution} where we plot the measured peak blue-shifted velocities as a function of time. It can also be seen in the left-hand panels of Figure~\ref{O_vel} which suggests a greater expansion velocity of the O-rich vs.\ the S, Ar-rich layers in the progenitor star. A 300 km s$^{-1}$ velocity difference is also seen in the maximum red-shifted FWZI of +4700 km s$^{-1}$  for  the [\ion{O}{2}] 7319,7330 emission line blend compared to the +4400 km s$^{-1}$ for the stronger and hence better measured [\ion{S}{3}] emission line at 9531 \AA. 
 
However, this O vs.\ S and Ar velocity difference appears a bit less, around 200 km s$^{-1}$, if considering the maximum blue-shifted expansion velocity of around $-4450$ km s$^{-1}$ for the oxygen emission lines versus $-4250$ km s$^{-1}$ for the S and Ar emission lines. In either case, in the 21.5 yr late-time spectrum of SN~1996cr sulfur and argon emission lines exhibit lower radial velocities than those of oxygen emission lines suggesting a velocity gradient in regard to ejecta abundances.

Oxygen and sulfur-rich ejecta emission from the blueshifted near side are several times the strength of the emissions from the redshifted rear ejecta (see Table~\ref{table_1}, Figs.~\ref{opt_spectrum} and \ref{O_vel}). This extinction effect is most obvious in the [\ion{S}{3}] $\lambda\lambda$9069,9531 line profiles where the redshifted components are only $\simeq$25\% the strength of the blueshifted component  (Fig.~\ref{O_vel}), and in the [\ion{S}{2}] emission where only the blueshifted component is readily visible and noticeably offset from the narrow emission from the H~II region (Fig.~\ref{opt_spectrum}). The extinction also leads to a near absence of redshifted [\ion{O}{3}] $\lambda\lambda$4959,5007 emission.

Observed and extinction corrected emission flux values along with estimated line luminosities are given in Table~\ref{table_1}. Based on a measured $N_{\rm H} = 0.81\pm 0.11$ $\times 10^{22}$ cm$^{-2}$ from 2018 {\sl Chandra} data (see Section 4.2) and using the recent formula of $N_{\rm H} = 2.87\pm 0.12 \times 10^{21}$ $A_{\rm V}$ cm$^{-2}$ \citep{Foight2016}, we estimate $A_{\rm V}$ = 2.82 mag. Adopting a value of 3.1 for R =  $A_{\rm V}$/$E(B-V)$ implying $E(B-V)$ = 0.91, we corrected by extinction the observed line fluxes. These values were used to estimate line luminosities assuming a distance of 4.2 Mpc.

\subsubsection{Broad and Narrow H$\alpha$ Emissions}


Along with the broad emissions from O, S, and Ar, broad and much weaker H$\alpha$ emission was also detected. This can been seen in Figure~\ref{fig:H_vel} where broad, weak H$\alpha$ emission can be seen at the base of the strong, narrow H II region lines of H$\alpha$ and [\ion{N}{2}]. Our modeling showed broad H$\alpha$ emission component extending approximately from $-2600$ to $+2600$ km s$^{-1}$ and slightly blueshifted, centered around $-220$ km s$^{-1}$. Enlarged H$\alpha$ and [\ion{N}{2}] line profiles are shown in Figure~\ref{fig:H_vel}.

Measured line widths of the narrow H~II region H$\alpha$, [\ion{N}{2}] $\lambda$6583, and [\ion{S}{2}] $\lambda$6716 lines was also found to be slight greater than that of the IMACS instrumental resolution as judged from the comparison neon arc lamp lines, that is 5.8 \AA \ vs.\ 5.6 \AA,  suggesting an intrinsic FWHM $\simeq$ 45 to 50 km  s$^{-1}$ for these emissions. This could imply either a real velocity dispersion due to one or more H~II regions seen toward or near SN~1996cr, or the expansion velocity of a single dominate H II region possibly associated with the SN.

The 2017 extinction corrected flux for the broad H$\alpha$ emission of $0.7 \times 10^{-14}$ is nearly 10 times less than \citet{Bauer2008} found in his 2006 spectrum. The FWHM of the line in 2017 ($\approx$1000 km s$^{-1}$) is about four times less than their 2006 value. The smoothness of the H$\alpha$ profile is an indication of the large number of line-emitting CSM clouds heated through interaction with the forward shock \citep{Chugai2021}.


\subsubsection{Deblending of Emission Components}

The broad oxygen, sulfur and argon emission features were deblended into two or three components assuming Gaussian profiles. This included blue- and red-shifted components, plus a narrow component where visible due to coincident H~II region emission. Because the [\ion{O}{2}] emission blend at $\lambda$7325 was the cleanest of the oxygen emission features, we used its profile as a template to determine the velocity centroid and width of the broad blueshifted and redshifted components for the other oxygen lines. In addition, for [\ion{O}{3}] $\lambda\lambda$4959,5007 lines, we forced the model flux ratios to match the intrinsic nebular 3:1 ratio for both broad and narrow components.

Modeled emission line strengths from this analysis are shown in the series of right-hand panels of Figure~\ref{O_vel}. Total blended emissions are shown as brown dashed lines, while the deconvolved red and blue shifted ejecta emissions are shown as a dotted blue lines. Narrow H~II region emission components are displayed as dotted red lines.

For the oxygen lines, our deblending model results indicate a redshifted component that is weaker but broader than the blueshifted component (FWHM = 2050 vs.\ 1220 km s$^{-1}$) while also displaying a much a lower peak velocity of only around $+1200$ km s$^{-1}$ compared to $-2330$ km s$^{-1}$ for the blueshifted component (see Table~\ref{Table_deblend}). In contrast, both the argon and sulfur lines showed more symmetric red and blueshifted peak velocities, with values around $\pm2000$ km s$^{-1}$. However, much like that seen for the oxygen lines, redshifted sulfur and argon components appear broader than their blueshifted components; e.g., FWHM = 2900 km s$^{-1}$ vs.\ 2500 km s$^{-1}$ for the [\ion{S}{3}] $\lambda$9531 line.



\begin{deluxetable}{lcccc}
\tablecaption{Blueshifted SN~1996cr Emission Peak Velocities at Ages 10, 21.5, and 25.5 Years \label{table:blueshifted}}
\tablewidth{0pc}
\tablehead{
\colhead{Line}  & \colhead{V$_{\rm 10 yr}$} &  \colhead{V$_{\rm 21 yr}$} & \colhead{V$_{\rm 25 yr}$} &
              \colhead{$\delta$(10:21, 21:25)}   \\ 
\colhead{ID}    & \colhead{km/s}       & \colhead{km/s}         & \colhead{km/s}        &
                   \colhead{km/s}   
}
\startdata
 $[$O I$]$ ~~~~6300  & $-3250$   & $-2380$ & $-1900$ & +870, +480 \\
 $[$O II$]$ ~~ 7325  & $-3200$   & $-2250$ & $-1800$ & +950, +450 \\
 $[$O III$]$  ~ 5007 & $-3250$    & $-2270$ & $-1800$ & +980, +470 \\
 $[$S III$]$ ~~ 9531  & \nodata  & $-1950$ & $-1550$ & \nodata, +400 \\
 $[$Ar III$]$ ~7136   & $-2800$  & $-1950$ & $-1600$ &  +850, +350 \\
 \enddata
\tablecomments{Velocity value for [\ion{O}{3}] at 10 yr taken from \citet{Bauer2008}.}
\end{deluxetable}

\begin{figure*}
\centering
\includegraphics[width=0.9\textwidth]{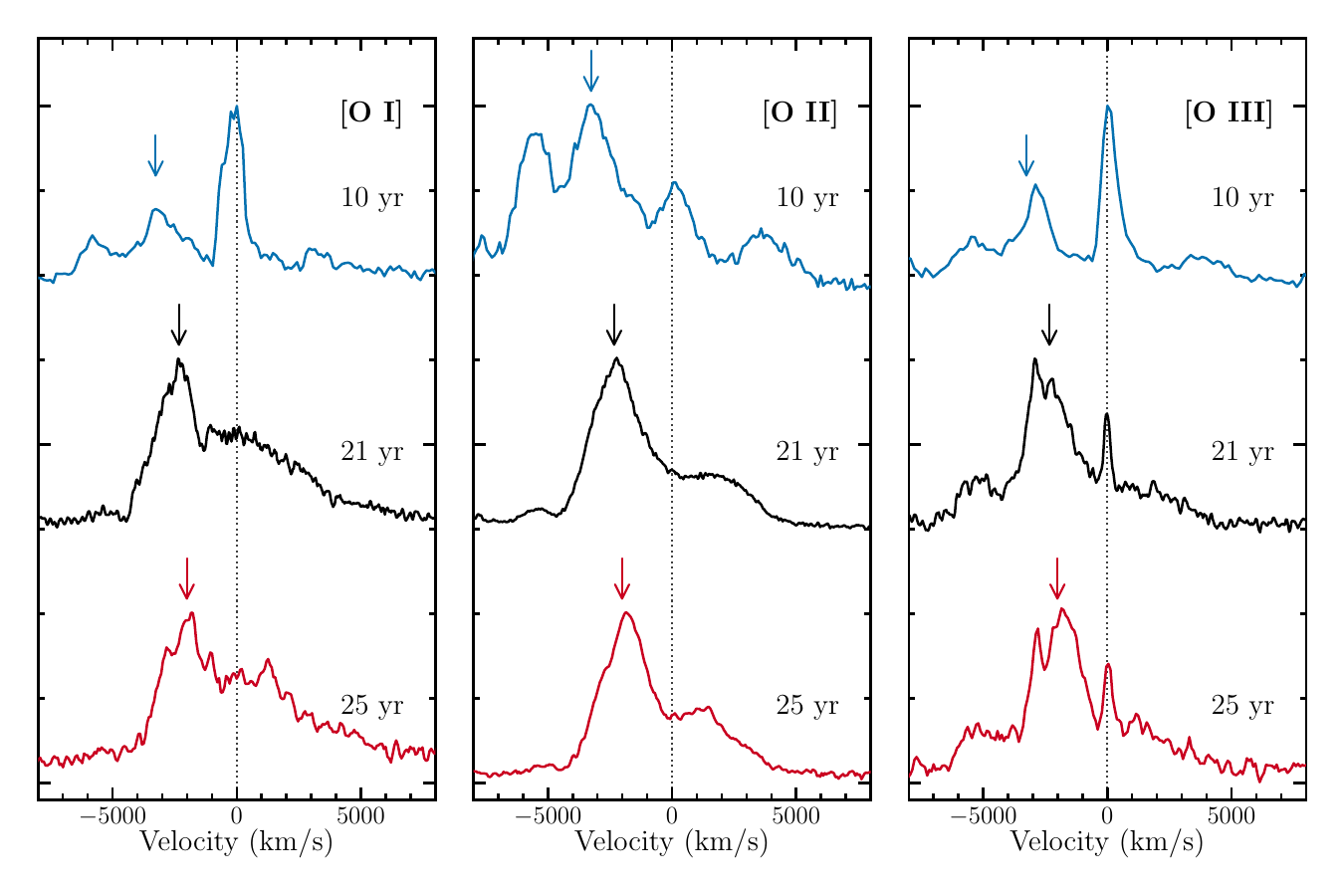}
\caption{Evolution of oxygen emission features at ages 10, 21.5, and 25.5 years. The arrows mark the approximate blueshifted velocities of the main emission peaks, indicating a decrease in peak emission velocity over the age span of 10 to 25 years. Broad blueshifted emission features dominates these oxygen emission profiles at years 21 and 25 (black and blue plots) compared to year 10 (red plot). Note: The [\ion{O}{2}] emission observed in the 10 year spectrum likely has some contribution from [\ion{Ca}{2}] emission which overlaps with blueshifted [\ion{O}{2}] emission. 
\label{O_evol}}
\end{figure*}

\begin{figure*}
    \centering
    \includegraphics[width=0.9\textwidth]{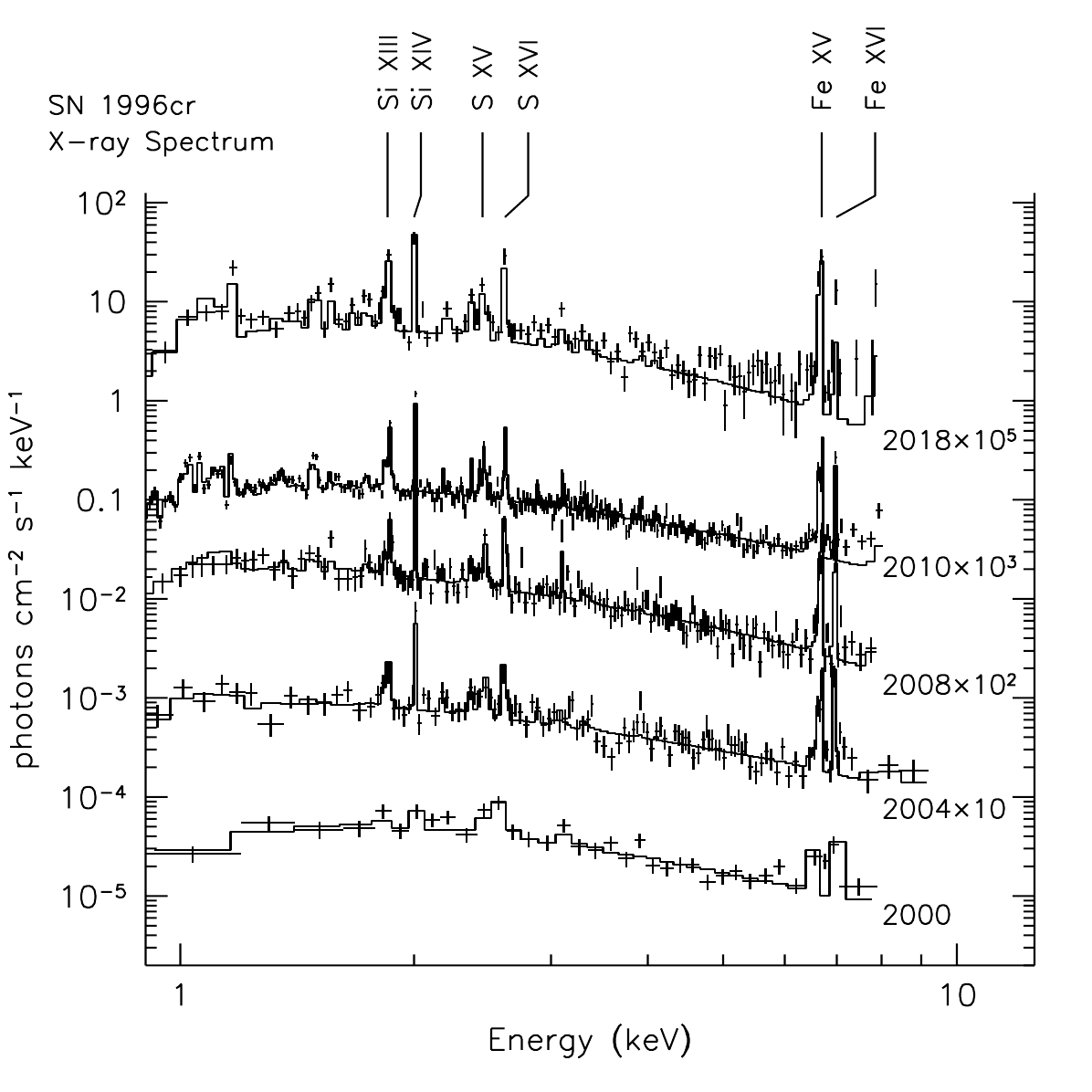}
    \caption{Unfolded 2000 to 2018 {\it Chandra} ACIS-S X-ray spectra. Relevant He- and H-like emission lines of Si, S, and Fe are shown. For clarity of comparison, spectra are displayed unfolded and arbitrarily scaled.}
    \label{fig:xray-spec}
\end{figure*}

\subsubsection{Evolution of Oxygen Line Emissions}

The evolution of SN~1996cr's oxygen emission lines between 10 and 25.5 years is presented in Figure~\ref{O_evol} where we overplot emission line profiles for the three ionization lines of oxygen; namely [\ion{O}{1}] 6300, 6364 \AA, [\ion{O}{2}] 7320,7330 \AA, and [\ion{O}{3}] 4959, 5007 \AA. Note that the discordant profile of [\ion{O}{2}] compared with the other oxygen line profiles is likely due to the presence of strong [\ion{Ca}{2}] 7291, 7324 \AA \ emission in the 10 yr spectrum which is largely absent in the 21 and 25 yr spectra. 

Late-time optical spectra of SN\,1996cr obtained with the VLT and spanning 2006-2019 were recently investigated by \citet{Niculescu-Duvaz2022}. These data are consistent with our own spectra, showing that the spectrum is dominated by ejecta emission. The forbidden oxygen emission line profiles were modeled with the Monte Carlo radiative transfer code DAMOCLES \citep{Bevan2017}, which quantifies the amount of newly formed dust causing absorption and scattering of SN ejecta line emission. Assuming a spherically symmetric distribution of ejecta, they are able to reproduce the emission line profiles including the blueshift of the oxygen line profiles (see Figure~\ref{O_vel}) by introducing   $\approx 0.1$\,M$_{\odot}$ of dust within the ejecta, having a high proportion of silicates (50-100\%) and grain radius 0.11 $\mu$m.

\subsubsection{Evolution Between Years 21.5 and 25.5}

While the flux level of the 2021 spectrum is about half that seen four years earlier, the 2017 and 2021 spectra appear remarkably similar, with only small differences seen in the emission profiles and/or relative strengths of some emission lines. 
Differences include the line profile of the [\ion{O}{1}] emission which is more strongly peaked in 2017 than in 2021. Also, although the broad [\ion{S}{3}] 9069 line appears to extend nearly 100 \AA \ further to the red in the 2021 spectrum, no such velocity extension is seen for the stronger [\ion{S}{3}] $\lambda$9531 lines. We do not have a good explanation for added emission around 9250 \AA. Weak but broad emission seen around H$\alpha$ is nearly symmetrical in the 2017 spectrum, but appears stronger on the blue-shifted side in the 2021 spectrum. However, this shift is not apparent in the ESO 2019 XShooter spectrum. A faint continuum was detected at both epochs, along with weak and very broad emission around 8600 \AA \ likely due to a blend line emissions of the \ion{Ca}{2} infrared triplet at 8498, 8542, 8662 \AA.

We also find differences in the velocities of the emission peaks for the oxygen, sulfur and argon lines between the 2017 and 2021 spectra. As shown in Table~2, the velocities of the blueshifted emission peaks for all O, S, and Ar emission lines decline by approximately 400 - 450 km s$^{-1}$. Greater decreases from large negative peak velocities with increasing age is seen between the 2006 spectrum and our 2017 spectrum (Table~\ref{table:blueshifted}), and such changes are increasing with time as shown in Figure~\ref{fig:oxy_evolution}.

\subsection{SN~1996cr's X-Ray Evolution}
\label{sec:cxcxray}

The \textit{Chandra} spectra taken from 2000 to 2018 are shown in Figure~\ref{fig:xray-spec}. To better discern changes in the X-ray emission, we present unfolded X-ray spectra from each epoch. These unfolded spectra allow us to better see changes in emission line strengths as predicted by the best fit model, from one epoch to the next, and also discern interesting features which may otherwise be undetectable when folded through the telescope response.

The {\it Chandra} data were fit using an absorbed, single temperature plane-parallel shock model. We chose abundances consistent with those adopted by \citet{QV2019} and allowed the hydrogen column density ($n_H$), electron temperature ($k_{\mathrm{B}}T_e$), and ionization age ($n_e t$) to all vary. The ionization age is especially important since it informs us on how long the X-ray emitting gas has been shocked. 

As a separate exercise, we also fit the line centroid for the Fe-K line complex near $\sim$ 6.7 keV. The rationale behind fitting the Fe-K line separately is that the centroid and line luminosity is strongly coupled to the degree of circumstellar interaction in both supernovae and supernova remnants  \citep{Yamaguchi2014,Patnaude2015,Margutti2017,Jacovich21}, though recent analyses have called in the validity of this technique in some objects \citep{siegel21}. 

\begin{deluxetable*}{llccccc}[ht]
\tablecolumns{7}
\tablewidth{0pt}
\tablecaption{Evolutionary Properties of SN~1996cr X-ray Emission \label{tab:XRAY}}
\tablehead{
\colhead{Obs. ID} & \colhead{Obs. Date} & \colhead{L$_X$} &
\colhead{$kT$} & \colhead{$n_H$} & 
\colhead{E(Fe)} & \colhead{L$_{Fe-K}$} \\ 
\colhead{} & \colhead{} & \colhead{10$^{39}$ erg s$^{-1}$} &
\colhead{keV} & \colhead{10$^{22}$ cm$^{-2}$} & 
\colhead{keV} & \colhead{10$^{46}$ photon s$^{-1}$}
}
\startdata
62877 & 2000-06-16 &  2.2 & 12.8$\pm$2.5 & 0.89$\pm$0.10 & 6.74$\pm$0.11 & 2.8$\pm$1.1\\
4771 &  2004-11-28 &  4.1 & 24.9$\pm$8.3 & 0.77$\pm$0.07 & 6.72$\pm$0.05 & 6.5$\pm$1.1\\
10223 & 2008-12-15 &  3.8 & 9.5$\pm$1.3  & 0.69$\pm$0.07 & 6.68$\pm$0.06 & 5.0$\pm$0.4\\
12823 & 2010-12-17 &  3.1 & 16.9$\pm$1.3 & 0.84$\pm$0.05 & 6.71$\pm$0.05 & 4.5$\pm$0.3\\
21688 \& 21974 & 2018-11-30 & 1.2 & 7.9$\pm$0.8 & 0.81$\pm$0.11 & 6.70$\pm$0.03 & 3.3$\pm$0.3 \\
\enddata
\end{deluxetable*}

As discussed in \citet{QV2019}, a high-resolution {\sl Chandra} spectrum of SN1996cr suggests there are likely two plasma components giving rise to the total observed X-ray emission. Additionally, their fits to a deep HETG observation suggest a complicated geometry for two emitting components. In our shorter observations, the data quality are such that trying to model the emission with two different plasma components results in too many degrees of freedom  ($\chi^2_{\mathrm{red}}$ $<<$ 1). As we are looking to interpret changes from one epoch to the next, we chose the simplest models whose results would still allow us to connect changes in the fitted parameters to changes in the physical properties of the remnant. In reality,
the X-ray emission is a combination of multiple components \citep[][e.g.,]{QV2019}, but trying to fit the lower CCD resolution
data to a model that accounts for multiple emitting components could result in overfitting the parameters.


\subsubsection{X-ray Emission Analysis}

The results from our spectral fits to the \textit{Chandra} spectra are shown in Table~\ref{tab:XRAY}. As discussed elsewhere \citep[e.g.,][]{QV2019}, SN~1996cr showed a noticeable increase in flux around 2000 -- 2004 but has since steadily declined. This is consistent with our observations (see Table~\ref{tab:XRAY}). We note that our spectral fits do not correlate well with those of \citet{QV2019}. This is likely due to the simpler model we used for fitting the lower resolution CCD data. As seen in the {\sl Chandra} spectra, (Fig.~\ref{fig:xray-spec}), the SN emission at epoch 2000 was initially dominated by line emission from H-like ions of silicon, sulfur, and iron. When coupled with a near absence of strong He-like emission. The absence of strong line emission below $\sim$ 1.8 keV could be due to multiple reasons, including the low spectral resolution of the ACIS CCD, or possibly because of the high temperature of the emitting material. In any case, the absence of line emission below silicon suggests a rapid heating of circumstellar material via a strong shock, similar to that which is observed in SN~2014C \citep{Margutti2017}.

Subsequent epochs, namely from 2004 -- 2010, reveal an X-ray spectrum initially dominated by H-like emission from silicon, sulfur, and iron. The L(H-like)/L(He-like) ratio  $>$ 1 is temperature sensitive and is consistent with the high temperatures found in the spectral fits. By 2018, the ratios of the H- to He-like states are $\approx$ 1, and $<$ 1 for \ion{Fe}{25} and \ion{Fe}{26} and we interpret this as due to recombination as the shock continues to decelerate and the material behind it cools. Absent in the pre-2010 spectra (Figure~\ref{fig:xray-spec}), is strong iron emission below \ion{Fe}{25}, which does appear in the 2010 and 2018 spectra. This suggests that the emission around 6.4 keV could be fluorescence line emission from neutral iron interacting with the shock, but the quality of the data do not allow for a definitive interpretation. 

As already noted, during the time period 2004-2010, SN~1996cr's X-ray luminosity is observed to decline steadily. The modeled electron temperature, column density, and Fe-K line centroid were observed to initially drop but then rise in 2010. We interpret the rise in 2010 as a result of the supernova ejecta catching up with shocked circumstellar material. At the same time, strong emission is observed from He-like states of silicon, sulfur, and iron. By 2010, emission from lower Z elements such as magnesium become prominent, as well as blended neon and Fe-L emission around 1 keV though emission from these elements are detected in epoch 2004;][]{QV2019}. 

Of particular note is that during the 2004--2010 observations emission from H-like ions is still higher or comparable to emission from He-like ions, indicating that the shocked gas remains highly ionized, and from epoch to epoch, variations in the relative intensities of the He- and H-like lines suggests rapid recombination as the shocked gas cools. This last point is supported by the onset of L-shell Fe emission. 

By the time of our {\sl Chandra} 2018 observation, the X-ray spectrum of SN~1996cr evolved such that \ion{Fe}{25} now dominates over \ion{Fe}{26}, and H- and He-like emission from silicon and sulfur are now comparable. As discussed in \citet{QV2019}, the shock may have recently broken free from any circumstellar shell. Interestingly, as seen in Figure~\ref{fig:xray-spec}, there appears to be excess X-ray emission above 7 keV. This emission first appears in 2010 and may represent the onset of radiative recombination continua (RRC) from highly ionized iron. RRC is commonly seen in Galactic SNR \citep[e.g., SNR W49B][]{Yamaguchi2018}, but its origin, due either strong interaction between a shock and a dense molecular cloud or a rapid adiabatic expansion caused by the breakout of a shock from a dense to a rarified ISM region, remains unclear \citep[e.g.][]{Moriya2012}. In any case, further observations may reveal the emergence of additional RRC from lower Z elements, as the SNR evolves.

\section{Discussion}

Below we discuss our recent X-ray and optical data with that of earlier data and then note similarities of SN~1996cr's optical spectrum with young O-rich remnants, and propose a scenario for its late-time evolutionary path for SN~1996cr making connections between its optical and X-ray kinematics and emission properties. We begin with an overview of optical spectra changes that occurred roughly one and two decades after outburst and compare its current spectrum to than of O-rich SNRs, followed by discussing its X-ray flux evolution and the progenitor's circumstellar Environment

\subsection{SN 1996cr's Late-Time Optical Evolution}

The dramatic difference between SN~1996cr's 10 year and 21 year spectra shown in Figure~\ref{Bauer_comp} and in  Table~\ref{Table_deblend} highlights the evolution of this young SN remnant as it transitioned from a H-rich envelope + CSM emission source to an ejecta dominated emission phase \citep{Mili2017}. This transition observed in the optical seems consistent with the trend observed in the X-ray \citep{QV2019}. Its fairly steady optical spectrum seen in its 21 and 25 yr old spectra (Fig.\ 4) may signal that this transition has moved into a much slower evolutionary phase.

The drivers for this transition are SN–CSM interactions which lead to the formation of a strong reverse shock moving back into the expanding ejecta thereby lighting up inner ejecta leading to late-time optical lines from intermediate-mass elements (IME) such as oxygen, argon, and sulfur. Forbidden emission lines from these elements are the ones that dominate the late-time optical spectra of CCSN like SN~1996cr.

The timescale of this spectral evolution during the SN-to-SNR transition is dependent on the dynamics of the explosion and the mass and density of the local CSM/ISM environment that the expanding blast wave and ejecta are running into. For many SNe it can be as short as 1 to 2 years. However, it may be as short as $\sim100$ days (e.g., SN\,2014C; \citealt{Mili2015}), or as long as a decade (e.g., SN\,2004dk; \citealt{Mauerhan2018}).

\subsection{SN~1996cr's Late-Time X-ray Evolution: Clues About the Progenitor's Circumstellar Environment}

\begin{figure}
    \centering
    \includegraphics[width=0.9\columnwidth]{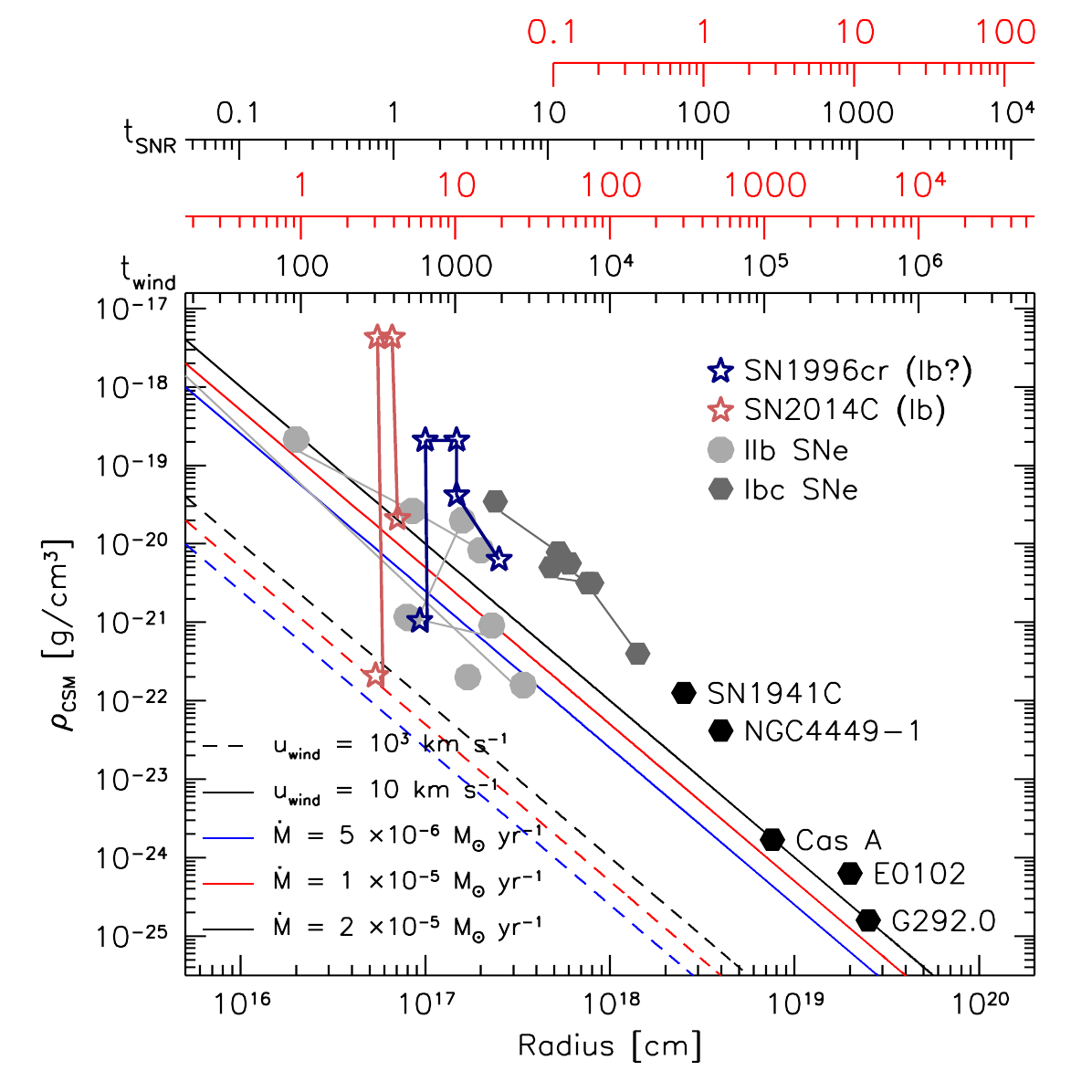}
    \caption{Circumstellar environments of several recent X-ray bright supernovae and supernova remnants. The X axis represents the radius of the forward shock either inferred by assuming homologous expansion of the supernova shock, or via direct measurement, in the case of nearby supernova remnants. The Y axis is the circumstellar density. The dashed lines represent $\rho_\mathrm{amb}(r)$ assuming isotropic mass loss, for the parameters listed in the legend. The top axes represent approximate ages for the SN, assuming homologous shock expansion with no deceleration into either fast ($v_\mathrm{wind}$ = 1000 km s$^{-1}$, red axis) or slow ($v_\mathrm{wind}$ = 10 km s$^{-1}$, black axis) winds, or the time before core collapse in the progenitor wind evolution which the shock is currently sampling ($t_\mathrm{SNR}$) for both fast (red) and slow (black) winds. Circumstellar densities for SN are inferred from the measured X-ray luminosity, as $L_X \propto (\dot{M}/v_\mathrm{wind})^2$. Data for Cas A, 1E 0102-7219, and G292.0+1.8 are taken from the literature \citep{lee10,lee14,xi19}.
    \label{fig:stripped-sne}}
\end{figure}

SN~1996cr was only serendipitously discovered several years after it is thought to have exploded \citep{Bauer2001}. As discussed elsewhere \citep{Bauer2008,Dewey2010,Dwark2012,QV2019} and confirmed in Table~\ref{tab:XRAY}, its X-ray emission was seen to increase for several years before declining between 2004 and 2008. Our 2018 observation is consistent with this trend, showing a continued decline from when SN~1996cr was last observed with {\it Chandra}. 

The evolution of the X-ray line emission, shown in Figure~\ref{fig:xray-spec}, is consistent with the picture described in \citet{Dwark2010} of a strong SN shock expanding in a low density environment before it is strongly decelerated by a dense circumstellar environment located $\sim$ 10$^{16}$ cm from the progenitor. We discuss this scenario in detail below.

\begin{figure*}
\centering
\includegraphics[width=0.9\textwidth]{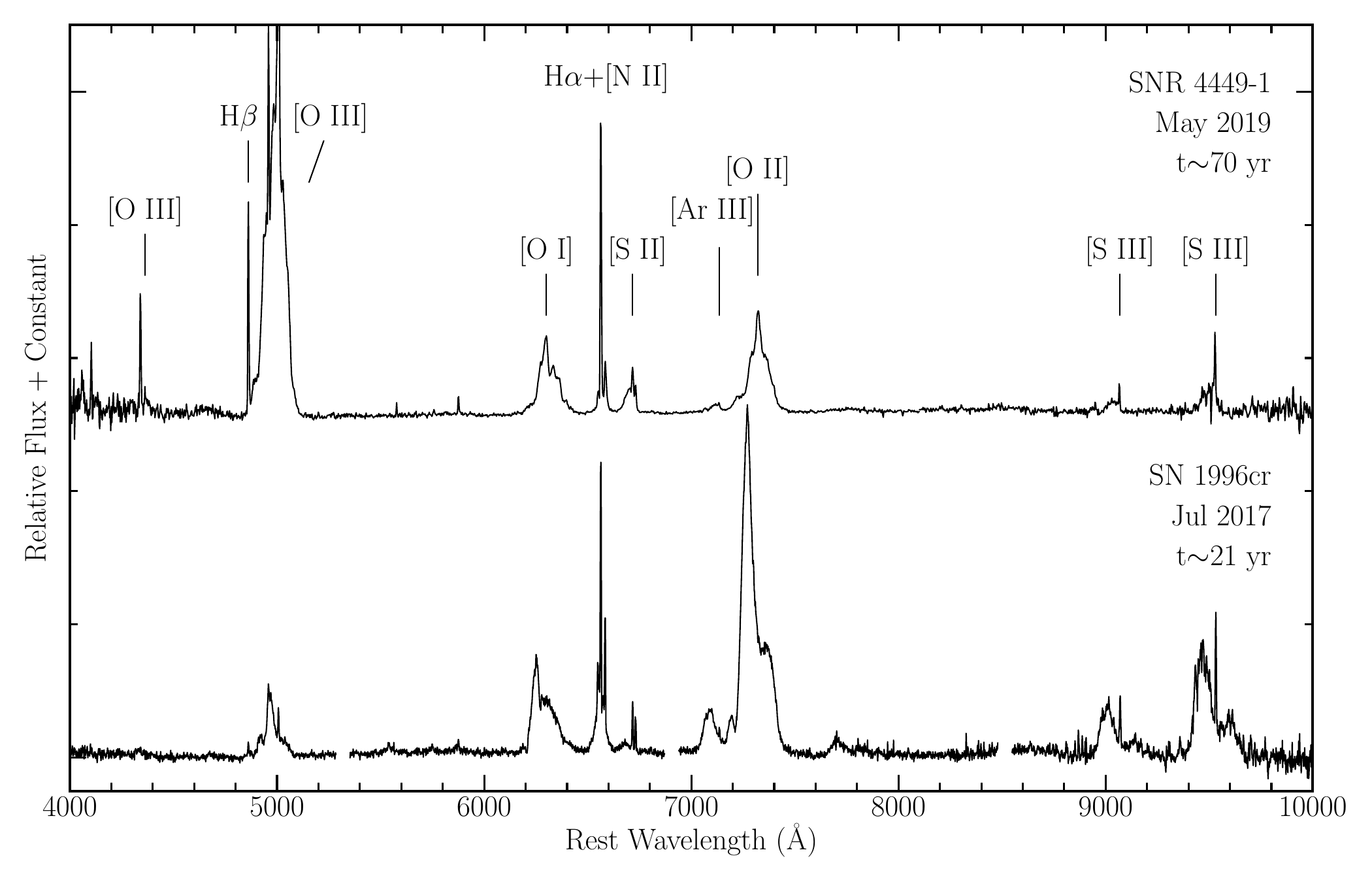}
\caption{A comparison the optical spectrum of SN~1996cr at an age of $\sim$ 21 years compared to the optical spectrum of SNR~4449-1 at an age of $\sim$ 70 years. Line identifications for the strongest emission features are marked.
\label{4449comp}}
\end{figure*}

The circumstellar profile we infer for SN~1996cr is similar to that of older objects, such as SN~1941C and Cas~A. As shown in Figure~\ref{fig:stripped-sne}, our latest observation suggests that the forward shock has now broken into the freely expanding wind of the red supergiant progenitor. The curves shown in Fig.~\ref{fig:stripped-sne} are for progenitor mass-loss rates up to 2$\times$10$^{-5}$ M$_{\sun}$ yr$^{-1}$, and our latest data point from 2018 suggests densities not too far from this value. When compared to observations of older SNe and young SNRs, the circumstellar properties of SN~1996cr are quite similar. 

Of particular note is the similarity between SN~1996cr and the O-rich SNR, Cas~A. Hydrodynamical models and bulk properties of the ejecta X-ray emission suggest that Cas A interacted with some sort of circumstellar cavity which formed during a late stage of the progenitor's evolution \citep{Patnaude2009,hwang09,weil20b,orlando22}. In particular, \citet{hwang09} showed that the ionization of the shocked ejecta is consistent with expansion in a low density cavity with a small radius ($r \approx 0.1$ pc), while \citet{orlando22} claim that the Cas A blastwave interacted with a circumstellar shell at a much larger radius ($r \approx 1.5$ pc), indicative of mass loss during a much earlier stage of the progenitor's evolution (10$^4$ -- 10$^{5}$ years before core collapse). The similarities of the inferred properties of objects like Cas A with the observed properties of much younger objects such as SN~1996cr suggests that mass loss events which lead to complicated circumstellar environments may be quite common. 

\subsection{Transition from a H-rich SN to an O-rich SNR}

The 2006 spectrum of SN~1996cr led  \citet{Bauer2007} to interpret SN~1996cr as a hydrogen-rich Type IIn. However, our 2017 and 2021 data show relatively little hydrogen emission, suggesting a SN with a relatively small mass of H-rich ejecta in its outer layers. 
Based on {\sl HST} and ground-based image, \citet{Bauer2008} concluded that $\sim$ 70\% of the observed H$\alpha$ emission was due to SN-CSM interaction, likely associated with photoionized wind material surrounding the progenitor. Now in its later stages of its evolution, SN~1996cr is better described as a young O-rich SNR rather than the H-rich SN event suggested by \citet{Bauer2008} based on spectra taken a decade earlier. As noted above, this has implications for the progenitor system, and suggests that the progenitor star was significantly stripped of its H-rich envelope at the time of its explosion.

\begin{deluxetable*}{lccccccccl}
\tablecolumns{10}
\tablewidth{0pt}
\tablecaption{Expansion Velocities and Luminosities of Young, O-Rich Core-Collapse SNRs
\label{O-rich-SNR}}
\label{fluxes}
\tablehead{\colhead{Supernova} & \colhead{Host}   & \colhead{Distance} & \colhead{Age}  & \colhead{V$_{\rm exp}$} &
 \multicolumn{3}{c}{\underline{~~~Lum. (10$^{36}$ erg s$^{-1}$)~~~}}  & \colhead{Obs.\ Epochs} & \colhead{References} \\
 \colhead{Remnant}     & \colhead{Galaxy} & \colhead{(Mpc)}    & \colhead{(yr)} & \colhead{(km s$^{-1}$)}  &
 \colhead{[O I]} & \colhead{ [O III] }   & \colhead{X-ray } & \colhead{opt:X-ray} & \colhead{ } }
\startdata
Cassiopeia A                  & Milky Way  & 0.0034   &  350   & $-4500$ to +6500   & 0.5    &  2.3  & 27     & 2015:2012       & 1, 2    \\
SN 1941C                      & NGC 4136   &  9.7     &   83        & $-2200$ to +4400   & 1.7    &  11   & 50     & 2019:2002  & 3, 4    \\
SNR 4449-1                    & NGC 4449   &  3.9     &   75        & $\pm5700$          & 26     &  145  & 240    & 2019:2001  & 5, 6, 7, 8  \\
SN 1957D                      & M83        &  4.6     &  67         & $\pm2700$          & 0.3    &  1.1  & 17     & 2011:2011  & 9, 10  \\
SN 1996cr                     &  Circinus   &  3.8    &  28         & $\pm4500$          & 11     & 5.4   & 4000   & 2017:2007  & 8, 10, 11    \\
\enddata
\tablenotetext{}{Note: Velocities correspond to [\ion{O}{1}] or [\ion{O}{3}] emissions. }
\tablenotetext{}{Note: Optical [\ion{O}{1}] 6300 + 6364 \AA, [\ion{O}{3}] 4959 + 5007 \AA, and X-ray luminosities are calculated
                  assuming the distance listed for the host galaxy. See \citet{Dwark2012} for tabulated X-ray energy range (keV) used to calculate X-ray luminosities. }
\tablenotetext{}{References: 1: \citet{Winkler2017};  2: \citet{Mili2013};
                             3: \cite{FW2020}; 4: \citet{Soria2008};
                             5: \citet{Mili2008}; 6: \citet{Patnaude2003}; 7: \citet{Summers2003}; 8: this paper; 
                             9: \citet{Long2012}; 10: \citet{Mili2012};
                             11: \citet{Bauer2008}
}

\end{deluxetable*}

In Figure~\ref{4449comp}, we show our 2017 optical spectrum of 1996cr compared to that of the young, and highly luminous O-rich SNR in NGC~4449. Broad oxygen emission lines dominate both SN spectra, with much of the line intensity differences to 96cr's high extinction plus SNR~4449-1's greater age ($\sim$ 70 yr vs.\ 21 yr) leading to a more evolved remnant with a greater amount of reverse shocked IME such as O and S. In Table~\ref{O-rich-SNR}, we compares SN~1996cr's expansion velocities and luminosities with that of several older O-rich SNRs. The optical emission expansion velocities and optical luminosities of these objects are not that different, with the exception of the SNR 4449-1 whose oxygen line emissions are an order of magnitude stronger than all the others.

We also note that SN~1996cr's observed late-time brightening is now recognized as being not that unusual. Numerous analog systems have been observed where stripped-envelope Type Ib and Ic supernovae experience late-time brightening consistent with interaction between the blast wave traveling beyond an interior bubble of low density material and encountering a higher density shell CSM. For example, at outburst SN~2014C was classified as a normal Type Ib, but optical spectra five months later showed that conspicuous H$\alpha$ emission with an overall FWHM velocity dispersion of approximately 1400~km\,s$^{-1}$ had emerged (Fig. 3). This is a feature associated with radiative shocks in dense clumps of CSM normally seen only in Type IIn SNe.  

This metamorphosis is consistent with a delayed interaction between an H-poor star's supernova explosion and a massive H-rich envelope that had been stripped decades to centuries before core collapse. In the case of SN~2001em, strong radio, X-ray, and H$\alpha$ emission was observed at an age of  2.5 yr \citep{CC2006}. The circumstellar shell was presumably formed by vigorous mass loss with a rate of  $(2-10) \times 10^{-3}\,M_{\odot}$ yr$^{-1}$ at $(1-2) \times 10^{3}$ yr prior to the supernova explosion. The hydrogen envelope was completely lost and subsequently was swept up and accelerated by the fast wind of the presupernova star up to a velocity of 30–50 km/s. 

SN~1996cr shares many properties with SN~2001em and SN~2014C, both of which exploded as stripped-envelope supernovae but later interacted with H-rich mass loss environments changing their spectroscopic properties from Type Ib/c to IIn \citep{chugai06,Margutti2017}. Applying a similar model to SN~1996cr suggests that the star underwent a major mass-losing event up to $\sim$ 1000 years before core collapse. As noted in \citet{Margutti2017}, this timescale is consistent with envelope ejection in binary systems or possibly due to convective instabilities during core C-burning \citep[e.g.,][]{Podsiadlowski92,arnett11}. 

Prominent intermediate-width H$\alpha$ emission was detected in the Type Ib explosion SN2004dk 4684 days (13 years) after discovery. Presumably, the SN blast wave had caught up with the hydrogen-rich CSM lost by the progenitor system. The data indicate that the mode of pre-SN mass loss was a relatively slow dense wind that persisted millennia before the SN, followed by a short-lived Wolf-Rayet phase that preceded core- collapse and created a cavity within an extended distribution of CSM. 

The physical mechanisms and time scales over which stripped-envelope SNe progenitor stars release their hydrogen envelopes are uncertain. Observations of Type Ib and Ic SNe provide opportunities to probe the origin of the necessary pre-SN mass loss through multi-wavelength signatures of interaction with distant CSM. Many comparisons have been made between SN\,1996cr and SN\,1987A. However, our data indicate that they come from very different progenitor stars. Whereas the progenitor of SN 1987A was an H-rich B3 I blue supergiant \citep{Podsiadlowski2017}, and the subsequent supernova still exhibits H-rich ejecta emission more than 25 years after explosion, the progenitor of SN\,1996cr was stripped of its hydrogen envelope, and any hydrogen emission is not associated with ejecta but with SN-CSM interaction.

\section{Conclusions}

We have presented new optical and X-ray observations of the young supernova remnant SN~1996cr in the nearby Circinus Galaxy. A summary of our main results and finding includes the following:

\begin{enumerate}
\item  We find dramatic difference between SN 1996cr’s 10 year and 21 year spectra, highlighing the evolution of this young SN remnant as it transitioned from a H-rich envelope + CSM emission source to an ejecta dominated emission phase. 

\item A fairly steady optical spectrum seen in 21 and 25 yr old spectra may signal that this transition has moved into a much slower evolutionary phase where strong reverse shock has begun interacting with expanding inner, more O-rich ejecta.

\item SN~1996cr's late-time spectrum now  resembles that seen in several young O-rich SNRs.

\item The latest X-ray observation suggests that SN~1996cr's forward shock has now broken into the freely expanding wind of the red supergiant progenitor. 
\end{enumerate}

Considering 1996cr's younger age compared to other O-rich extragalactic SN/SNRs (Table 5), continued monitoring of SN~199cr's optical and X-ray spectral and flux properties may offer better insight in understanding of late-time evolution of stripped envelope Type Ib/IIb SNe.


\acknowledgements 

D.~J.~P. acknowledges support from the {\sl Chandra} X-ray Center, which is operated by the Smithsonian Institution under NASA contract NAS8-03060. D.~M.\ acknowledges NSF support from grants PHY-1914448 and AST-2037297. D.~J.~P. acknowledges the \textit{Chandra} High Resolution Camera Guaranteed Time Program. D.M.\ acknowledges NSF support from grants PHY-2209451 and AST-2206532. Observations in this paper made use of this program. This paper employs a list of Chandra datasets, obtained by the Chandra X-ray Observatory, contained in~\dataset[DOI: 284]{https://doi.org/10.25574/cdc.284}

\bibliography{references}{}

\begin{thebibliography}{}
\expandafter\ifx\csname natexlab\endcsname\relax\def\natexlab#1{#1}\fi
\providecommand{\url}[1]{\href{#1}{#1}}
\providecommand{\dodoi}[1]{doi:~\href{http://doi.org/#1}{\nolinkurl{#1}}}
\providecommand{\doeprint}[1]{\href{http://ascl.net/#1}{\nolinkurl{http://ascl.net/#1}}}
\providecommand{\doarXiv}[1]{\href{https://arxiv.org/abs/#1}{\nolinkurl{https://arxiv.org/abs/#1}}}

\bibitem[{{Andrews} {et~al.}(2010){Andrews}, {Gallagher}, {Clayton},
  {Sugerman}, {Chatelain}, {Clem}, {Welch}, {Barlow}, {Ercolano}, {Fabbri},
  {Wesson}, \& {Meixner}}]{Andrews2010}
{Andrews}, J.~E., {Gallagher}, J.~S., {Clayton}, G.~C., {et~al.} 2010, \apj,
  715, 541, \dodoi{10.1088/0004-637X/715/1/541}

\bibitem[{{Andrews} {et~al.}(2011){Andrews}, {Sugerman}, {Clayton},
  {Gallagher}, {Barlow}, {Clem}, {Ercolano}, {Fabbri}, {Meixner}, {Otsuka},
  {Welch}, \& {Wesson}}]{Andrews2011}
{Andrews}, J.~E., {Sugerman}, B.~E.~K., {Clayton}, G.~C., {et~al.} 2011, \apj,
  731, 47, \dodoi{10.1088/0004-637X/731/1/47}

\bibitem[{{Arnett} \& {Meakin}(2011)}]{arnett11}
{Arnett}, W.~D., \& {Meakin}, C. 2011, \apj, 741, 33,
  \dodoi{10.1088/0004-637X/741/1/33}

\bibitem[{{Bauer}(2007)}]{Bauer2007}
{Bauer}, F. 2007, Central Bureau Electronic Telegrams, 879, 14

\bibitem[{{Bauer} {et~al.}(2001){Bauer}, {Brandt}, {Sambruna}, {Chartas},
  {Garmire}, {Kaspi}, \& {Netzer}}]{Bauer2001}
{Bauer}, F.~E., {Brandt}, W.~N., {Sambruna}, R.~M., {et~al.} 2001, \aj, 122,
  182, \dodoi{10.1086/321123}

\bibitem[{{Bauer} {et~al.}(2008){Bauer}, {Dwarkadas}, {Brandt}, {Immler},
  {Smartt}, {Bartel}, \& {Bietenholz}}]{Bauer2008}
{Bauer}, F.~E., {Dwarkadas}, V.~V., {Brandt}, W.~N., {et~al.} 2008, \apj, 688,
  1210, \dodoi{10.1086/589761}

\bibitem[{{Bevan} {et~al.}(2017){Bevan}, {Barlow}, \&
  {Milisavljevic}}]{Bevan2017}
{Bevan}, A., {Barlow}, M.~J., \& {Milisavljevic}, D. 2017, \mnras, 465, 4044,
  \dodoi{10.1093/mnras/stw2985}

\bibitem[{{Bietenholz} {et~al.}(2010){Bietenholz}, {Bartel}, {Milisavljevic},
  {Fesen}, {Challis}, \& {Kirshner}}]{Bietenholz2010}
{Bietenholz}, M.~F., {Bartel}, N., {Milisavljevic}, D., {et~al.} 2010, \mnras,
  409, 1594, \dodoi{10.1111/j.1365-2966.2010.17402.x}

\bibitem[{{Blair} {et~al.}(1983){Blair}, {Kirshner}, \& {Winkler}}]{Blair1983}
{Blair}, W.~P., {Kirshner}, R.~P., \& {Winkler}, P.~F., J. 1983, \apj, 272, 84,
  \dodoi{10.1086/161263}

\bibitem[{{Blair} {et~al.}(2015){Blair}, {Winkler}, {Long}, {Whitmore}, {Kim},
  {Soria}, {Kuntz}, {Plucinsky}, {Dopita}, \& {Stockdale}}]{Blair2015}
{Blair}, W.~P., {Winkler}, P.~F., {Long}, K.~S., {et~al.} 2015, \apj, 800, 118,
  \dodoi{10.1088/0004-637X/800/2/118}

\bibitem[{{Chevalier}(1982)}]{Chevalier1982}
{Chevalier}, R.~A. 1982, \apj, 258, 790, \dodoi{10.1086/160126}

\bibitem[{{Chevalier} \& {Fransson}(1994)}]{CF94}
{Chevalier}, R.~A., \& {Fransson}, C. 1994, \apj, 420, 268,
  \dodoi{10.1086/173557}

\bibitem[{{Chevalier} \& {Fransson}(2006)}]{chevalier06}
---. 2006, \apj, 651, 381, \dodoi{10.1086/507606}

\bibitem[{{Chevalier} \& {Fransson}(2017)}]{CF2017}
---. 2017, {Thermal and Non-thermal Emission from Circumstellar Interaction},
  ed. A.~W. {Alsabti} \& P.~{Murdin}, 875,
  \dodoi{10.1007/978-3-319-21846-5\_34}

\bibitem[{{Chugai}(2021)}]{Chugai2021}
{Chugai}, N.~N. 2021, \mnras, 508, 6023, \dodoi{10.1093/mnras/stab2981}

\bibitem[{{Chugai} \& {Chevalier}(2006{\natexlab{a}})}]{CC2006}
{Chugai}, N.~N., \& {Chevalier}, R.~A. 2006{\natexlab{a}}, \apj, 641, 1051,
  \dodoi{10.1086/500539}

\bibitem[{{Chugai} \& {Chevalier}(2006{\natexlab{b}})}]{chugai06}
---. 2006{\natexlab{b}}, \apj, 641, 1051, \dodoi{10.1086/500539}

\bibitem[{{Chugai} {et~al.}(2004){Chugai}, {Blinnikov}, {Cumming}, {Lundqvist},
  {Bragaglia}, {Filippenko}, {Leonard}, {Matheson}, \&
  {Sollerman}}]{Chugai2004}
{Chugai}, N.~N., {Blinnikov}, S.~I., {Cumming}, R.~J., {et~al.} 2004, \mnras,
  352, 1213, \dodoi{10.1111/j.1365-2966.2004.08011.x}

\bibitem[{{Cowan} {et~al.}(1994){Cowan}, {Roberts}, \& {Branch}}]{Cowan1994}
{Cowan}, J.~J., {Roberts}, D.~A., \& {Branch}, D. 1994, \apj, 434, 128,
  \dodoi{10.1086/174710}

\bibitem[{{Dessart} {et~al.}(2015){Dessart}, {Audit}, \&
  {Hillier}}]{Dessart2015}
{Dessart}, L., {Audit}, E., \& {Hillier}, D.~J. 2015, \mnras, 449, 4304,
  \dodoi{10.1093/mnras/stv609}

\bibitem[{{Dewey} {et~al.}(2011){Dewey}, {Bauer}, \& {Dwarkadas}}]{Dewey2011}
{Dewey}, D., {Bauer}, F.~E., \& {Dwarkadas}, V.~V. 2011, in American Institute
  of Physics Conference Series, Vol. 1358, American Institute of Physics
  Conference Series, ed. J.~E. {McEnery}, J.~L. {Racusin}, \& N.~{Gehrels},
  289--292, \dodoi{10.1063/1.3621791}

\bibitem[{{Dwarkadas} {et~al.}(2010{\natexlab{a}}){Dwarkadas}, {Dewey}, \&
  {Bauer}}]{Dewey2010}
{Dwarkadas}, V.~V., {Dewey}, D., \& {Bauer}, F. 2010{\natexlab{a}}, \mnras,
  407, 812, \dodoi{10.1111/j.1365-2966.2010.16966.x}

\bibitem[{{Dwarkadas} {et~al.}(2010{\natexlab{b}}){Dwarkadas}, {Dewey}, \&
  {Bauer}}]{Dwark2010}
---. 2010{\natexlab{b}}, \mnras, 407, 812,
  \dodoi{10.1111/j.1365-2966.2010.16966.x}

\bibitem[{{Dwarkadas} \& {Gruszko}(2012)}]{Dwark2012}
{Dwarkadas}, V.~V., \& {Gruszko}, J. 2012, \mnras, 419, 1515,
  \dodoi{10.1111/j.1365-2966.2011.19808.x}

\bibitem[{{Eck} {et~al.}(2002){Eck}, {Cowan}, \& {Branch}}]{Eck2002}
{Eck}, C.~R., {Cowan}, J.~J., \& {Branch}, D. 2002, \apj, 573, 306,
  \dodoi{10.1086/340583}

\bibitem[{{Fesen} \& {Becker}(1990)}]{Fesen1990}
{Fesen}, R.~A., \& {Becker}, R.~H. 1990, \apj, 351, 437, \dodoi{10.1086/168480}

\bibitem[{{Fesen} \& {Matonick}(1993)}]{Fesen1993}
{Fesen}, R.~A., \& {Matonick}, D.~M. 1993, \apj, 407, 110,
  \dodoi{10.1086/172496}

\bibitem[{{Fesen} \& {Weil}(2020)}]{FW2020}
{Fesen}, R.~A., \& {Weil}, K.~E. 2020, \apj, 890, 15,
  \dodoi{10.3847/1538-4357/ab67b7}

\bibitem[{{Foight} {et~al.}(2016){Foight}, {G{\"u}ver}, {{\"O}zel}, \&
  {Slane}}]{Foight2016}
{Foight}, D.~R., {G{\"u}ver}, T., {{\"O}zel}, F., \& {Slane}, P.~O. 2016, \apj,
  826, 66, \dodoi{10.3847/0004-637X/826/1/66}

\bibitem[{{Fransson} {et~al.}(2002){Fransson}, {Chevalier}, {Filippenko},
  {Leibundgut}, {Barth}, {Fesen}, {Kirshner}, {Leonard}, {Li}, {Lundqvist},
  {Sollerman}, \& {Van Dyk}}]{Fransson2002}
{Fransson}, C., {Chevalier}, R.~A., {Filippenko}, A.~V., {et~al.} 2002, \apj,
  572, 350, \dodoi{10.1086/340295}

\bibitem[{{Fransson} {et~al.}(2014){Fransson}, {Ergon}, {Challis}, {Chevalier},
  {France}, {Kirshner}, {Marion}, {Milisavljevic}, {Smith}, {Bufano},
  {Friedman}, {Kangas}, {Larsson}, {Mattila}, {Benetti}, {Chornock}, {Czekala},
  {Soderberg}, \& {Sollerman}}]{Fransson2014}
{Fransson}, C., {Ergon}, M., {Challis}, P.~J., {et~al.} 2014, \apj, 797, 118,
  \dodoi{10.1088/0004-637X/797/2/118}

\bibitem[{{Freeman} {et~al.}(1977){Freeman}, {Karlsson}, {Lynga}, {Burrell},
  {van Woerden}, {Goss}, \& {Mebold}}]{Freeman1977}
{Freeman}, K.~C., {Karlsson}, B., {Lynga}, G., {et~al.} 1977, \aap, 55, 445

\bibitem[{{Gal-Yam} \& {Leonard}(2009)}]{Gal-Yam2009}
{Gal-Yam}, A., \& {Leonard}, D.~C. 2009, \nat, 458, 865,
  \dodoi{10.1038/nature07934}

\bibitem[{{Graham} {et~al.}(2015){Graham}, {Nugent}, {Sullivan}, {Filippenko},
  {Cenko}, {Silverman}, {Clubb}, \& {Zheng}}]{Graham2015}
{Graham}, M.~L., {Nugent}, P.~E., {Sullivan}, M., {et~al.} 2015, \mnras, 454,
  1948, \dodoi{10.1093/mnras/stv1888}

\bibitem[{{Hamuy} {et~al.}(1994){Hamuy}, {Suntzeff}, {Heathcote}, {Walker},
  {Gigoux}, \& {Phillips}}]{Hamuy1994}
{Hamuy}, M., {Suntzeff}, N.~B., {Heathcote}, S.~R., {et~al.} 1994, \pasp, 106,
  566, \dodoi{10.1086/133417}

\bibitem[{{Hamuy} {et~al.}(1992){Hamuy}, {Walker}, {Suntzeff}, {Gigoux},
  {Heathcote}, \& {Phillips}}]{Hamuy1992}
{Hamuy}, M., {Walker}, A.~R., {Suntzeff}, N.~B., {et~al.} 1992, \pasp, 104,
  533, \dodoi{10.1086/133028}

\bibitem[{{Hwang} \& {Laming}(2009)}]{hwang09}
{Hwang}, U., \& {Laming}, J.~M. 2009, \apj, 703, 883,
  \dodoi{10.1088/0004-637X/703/1/883}

\bibitem[{{Jacovich} {et~al.}(2021){Jacovich}, {Patnaude}, {Slane}, {Badenes},
  {Lee}, {Nagataki}, \& {Milisavljevic}}]{Jacovich21}
{Jacovich}, T., {Patnaude}, D., {Slane}, P., {et~al.} 2021, \apj, 914, 41,
  \dodoi{10.3847/1538-4357/abf935}

\bibitem[{{Jones}(1941)}]{Jones41}
{Jones}, R. 1941, \iaucirc, 866, 1

\bibitem[{{Kerzendorf} {et~al.}(2017){Kerzendorf}, {McCully}, {Taubenberger},
  {Jerkstrand}, {Seitenzahl}, {Ruiter}, {Spyromilio}, {Long}, \&
  {Fransson}}]{Kerz2017}
{Kerzendorf}, W.~E., {McCully}, C., {Taubenberger}, S., {et~al.} 2017, \mnras,
  472, 2534, \dodoi{10.1093/mnras/stx1923}

\bibitem[{{Kiewe} {et~al.}(2012){Kiewe}, {Gal-Yam}, {Arcavi}, {Leonard},
  {Emilio Enriquez}, {Cenko}, {Fox}, {Moon}, {Sand }, {Soderberg}, \&
  {CCCP}}]{Kiewe2012}
{Kiewe}, M., {Gal-Yam}, A., {Arcavi}, I., {et~al.} 2012, \apj, 744, 10,
  \dodoi{10.1088/0004-637X/744/1/10}

\bibitem[{{Kirshner} \& {Blair}(1980)}]{Kirshner1980}
{Kirshner}, R.~P., \& {Blair}, W.~P. 1980, \apj, 236, 135,
  \dodoi{10.1086/157726}

\bibitem[{{Koribalski} {et~al.}(2004){Koribalski}, {Staveley-Smith}, {Kilborn},
  {Ryder}, {Kraan-Korteweg}, {Ryan-Weber}, {Ekers}, {Jerjen}, {Henning},
  {Putman}, {Zwaan}, {de Blok}, {Calabretta}, {Disney}, {Minchin}, {Bhathal},
  {Boyce}, {Drinkwater}, {Freeman}, {Gibson}, {Green}, {Haynes}, {Juraszek},
  {Kesteven}, {Knezek}, {Mader}, {Marquarding}, {Meyer}, {Mould}, {Oosterloo},
  {O'Brien}, {Price}, {Sadler}, {Schr{\"o}der}, {Stewart}, {Stootman}, {Waugh},
  {Warren}, {Webster}, \& {Wright}}]{Korib2004}
{Koribalski}, B.~S., {Staveley-Smith}, L., {Kilborn}, V.~A., {et~al.} 2004,
  \aj, 128, 16, \dodoi{10.1086/421744}

\bibitem[{{Kundu} {et~al.}(2019){Kundu}, {Lundqvist}, {Sorokina},
  {P{\'e}rez-Torres}, {Blinnikov}, {O'Connor}, {Ergon}, {Chandra}, \&
  {Das}}]{Kundu2019}
{Kundu}, E., {Lundqvist}, P., {Sorokina}, E., {et~al.} 2019, \apj, 875, 17,
  \dodoi{10.3847/1538-4357/ab0d81}

\bibitem[{{Lee} {et~al.}(2014){Lee}, {Park}, {Hughes}, \& {Slane}}]{lee14}
{Lee}, J.-J., {Park}, S., {Hughes}, J.~P., \& {Slane}, P.~O. 2014, \apj, 789,
  7, \dodoi{10.1088/0004-637X/789/1/7}

\bibitem[{{Lee} {et~al.}(2010){Lee}, {Park}, {Hughes}, {Slane}, {Gaensler},
  {Ghavamian}, \& {Burrows}}]{lee10}
{Lee}, J.-J., {Park}, S., {Hughes}, J.~P., {et~al.} 2010, \apj, 711, 861,
  \dodoi{10.1088/0004-637X/711/2/861}

\bibitem[{{Lianou} {et~al.}(2019){Lianou}, {Barmby}, {Mosenkov}, {Lehnert}, \&
  {Karczewski}}]{Lianou2019}
{Lianou}, S., {Barmby}, P., {Mosenkov}, A.~A., {Lehnert}, M., \& {Karczewski},
  O. 2019, \aap, 631, A38, \dodoi{10.1051/0004-6361/201834553}

\bibitem[{{Long} {et~al.}(1989){Long}, {Blair}, \& {Krzeminski}}]{Long1989}
{Long}, K.~S., {Blair}, W.~P., \& {Krzeminski}, W. 1989, \apjl, 340, L25,
  \dodoi{10.1086/185430}

\bibitem[{{Long} {et~al.}(2019){Long}, {Winkler}, \& {Blair}}]{Long2019}
{Long}, K.~S., {Winkler}, P.~F., \& {Blair}, W.~P. 2019, \apj, 875, 85,
  \dodoi{10.3847/1538-4357/ab0d94}

\bibitem[{{Long} {et~al.}(2012){Long}, {Blair}, {Godfrey}, {Kuntz},
  {Plucinsky}, {Soria}, {Stockdale}, {Whitmore}, \& {Winkler}}]{Long2012}
{Long}, K.~S., {Blair}, W.~P., {Godfrey}, L.~E.~H., {et~al.} 2012, \apj, 756,
  18, \dodoi{10.1088/0004-637X/756/1/18}

\bibitem[{{Margutti} {et~al.}(2017){Margutti}, {Kamble}, {Milisavljevic},
  {Zapartas}, {de Mink}, {Drout}, {Chornock}, {Risaliti}, {Zauderer},
  {Bietenholz}, {Cantiello}, {Chakraborti}, {Chomiuk}, {Fong}, {Grefenstette},
  {Guidorzi}, {Kirshner}, {Parrent}, {Patnaude}, {Soderberg}, {Gehrels}, \&
  {Harrison}}]{Margutti2017}
{Margutti}, R., {Kamble}, A., {Milisavljevic}, D., {et~al.} 2017, \apj, 835,
  140, \dodoi{10.3847/1538-4357/835/2/140}

\bibitem[{{Martini} {et~al.}(2011){Martini}, {Stoll}, {Derwent}, {Zhelem},
  {Atwood}, {Gonzalez}, {Mason}, {O'Brien}, {Pappalardo}, \&
  {Pogge}}]{Martini2011}
{Martini}, P., {Stoll}, R., {Derwent}, M.~A., {et~al.} 2011, \pasp, 123, 187,
  \dodoi{10.1086/658357}

\bibitem[{{Massey} \& {Gronwall}(1990)}]{Massey90}
{Massey}, P., \& {Gronwall}, C. 1990, \apj, 358, 344, \dodoi{10.1086/168991}

\bibitem[{{Mauerhan} {et~al.}(2018){Mauerhan}, {Filippenko}, {Zheng}, {Brink},
  {Graham}, {Shivvers}, \& {Clubb}}]{Mauerhan2018}
{Mauerhan}, J.~C., {Filippenko}, A.~V., {Zheng}, W., {et~al.} 2018, \mnras,
  478, 5050, \dodoi{10.1093/mnras/sty1307}

\bibitem[{{McCray}(1993)}]{McCray1993}
{McCray}, R. 1993, \araa, 31, 175, \dodoi{10.1146/annurev.aa.31.090193.001135}

\bibitem[{{Meunier} {et~al.}(2013){Meunier}, {Bauer}, {Dwarkadas},
  {Koribalski}, {Emonts}, {Hunstead}, {Campbell-Wilson}, {Stockdale}, \&
  {Tingay}}]{Meunier2013}
{Meunier}, C., {Bauer}, F.~E., {Dwarkadas}, V.~V., {et~al.} 2013, \mnras, 431,
  2453, \dodoi{10.1093/mnras/stt340}

\bibitem[{{Milisavljevic} \& {Fesen}(2008)}]{Mili2008}
{Milisavljevic}, D., \& {Fesen}, R.~A. 2008, \apj, 677, 306,
  \dodoi{10.1086/528929}

\bibitem[{{Milisavljevic} \& {Fesen}(2013)}]{Mili2013}
---. 2013, \apj, 772, 134, \dodoi{10.1088/0004-637X/772/2/134}

\bibitem[{{Milisavljevic} \& {Fesen}(2017)}]{Mili2017}
---. 2017, {The Supernova - Supernova Remnant Connection}, ed. A.~W. {Alsabti}
  \& P.~{Murdin}, 2211, \dodoi{10.1007/978-3-319-21846-5_97}

\bibitem[{{Milisavljevic} {et~al.}(2012){Milisavljevic}, {Fesen}, {Chevalier},
  {Kirshner}, {Challis}, \& {Turatto}}]{Mili2012}
{Milisavljevic}, D., {Fesen}, R.~A., {Chevalier}, R.~A., {et~al.} 2012, \apj,
  751, 25, \dodoi{10.1088/0004-637X/751/1/25}

\bibitem[{{Milisavljevic} {et~al.}(2018){Milisavljevic}, {Patnaude},
  {Chevalier}, {Raymond}, {Fesen}, {Margutti}, {Conner}, \&
  {Banovetz}}]{Mili2018}
{Milisavljevic}, D., {Patnaude}, D.~J., {Chevalier}, R.~A., {et~al.} 2018,
  \apjl, 864, L36, \dodoi{10.3847/2041-8213/aadd4e}

\bibitem[{{Milisavljevic} {et~al.}(2015){Milisavljevic}, {Margutti}, {Kamble},
  {Patnaude}, {Raymond}, {Eldridge}, {Fong}, {Bietenholz}, {Challis},
  {Chornock}, {Drout}, {Fransson}, {Fesen}, {Grindlay}, {Kirshner}, {Lunnan},
  {Mackey}, {Miller}, {Parrent}, {Sand ers}, {Soderberg}, \&
  {Zauderer}}]{Mili2015}
{Milisavljevic}, D., {Margutti}, R., {Kamble}, A., {et~al.} 2015, \apj, 815,
  120, \dodoi{10.1088/0004-637X/815/2/120}

\bibitem[{{Moriya}(2012)}]{Moriya2012}
{Moriya}, T.~J. 2012, \apjl, 750, L13, \dodoi{10.1088/2041-8205/750/1/L13}

\bibitem[{{Niculescu-Duvaz} {et~al.}(2022){Niculescu-Duvaz}, {Barlow}, {Bevan},
  {Wesson}, {Milisavljevic}, {De Looze}, {Clayton}, {Krafton}, {Matsuura}, \&
  {Brady}}]{Niculescu-Duvaz2022}
{Niculescu-Duvaz}, M., {Barlow}, M.~J., {Bevan}, A., {et~al.} 2022, arXiv
  e-prints, arXiv:2204.14179.
\newblock \doarXiv{2204.14179}

\bibitem[{{Oke}(1974)}]{Oke74}
{Oke}, J.~B. 1974, \apjs, 27, 21, \dodoi{10.1086/190287}

\bibitem[{{Orlando} {et~al.}(2022){Orlando}, {Wongwathanarat}, {Janka},
  {Miceli}, {Nagataki}, {Ono}, {Bocchino}, {Vink}, {Milisavljevic}, {Patnaude},
  \& {Peres}}]{orlando22}
{Orlando}, S., {Wongwathanarat}, A., {Janka}, H.~T., {et~al.} 2022, \aap, 666,
  A2, \dodoi{10.1051/0004-6361/202243258}

\bibitem[{{Patnaude} \& {Fesen}(2003)}]{Patnaude2003}
{Patnaude}, D.~J., \& {Fesen}, R.~A. 2003, \apj, 587, 221,
  \dodoi{10.1086/368124}

\bibitem[{{Patnaude} \& {Fesen}(2009)}]{Patnaude2009}
---. 2009, \apj, 697, 535, \dodoi{10.1088/0004-637X/697/1/535}

\bibitem[{{Patnaude} {et~al.}(2015){Patnaude}, {Lee}, {Slane}, {Badenes},
  {Heger}, {Ellison}, \& {Nagataki}}]{Patnaude2015}
{Patnaude}, D.~J., {Lee}, S.-H., {Slane}, P.~O., {et~al.} 2015, \apj, 803, 101,
  \dodoi{10.1088/0004-637X/803/2/101}

\bibitem[{{Patnaude} {et~al.}(2017){Patnaude}, {Lee}, {Slane}, {Badenes},
  {Nagataki}, {Ellison}, \& {Milisavljevic}}]{Patnaude2017}
---. 2017, \apj, 849, 109, \dodoi{10.3847/1538-4357/aa9189}

\bibitem[{{Patnaude} {et~al.}(2011){Patnaude}, {Loeb}, \&
  {Jones}}]{Patnaude2011}
{Patnaude}, D.~J., {Loeb}, A., \& {Jones}, C. 2011, \na, 16, 187,
  \dodoi{10.1016/j.newast.2010.09.004}

\bibitem[{{Podsiadlowski}(2017)}]{Podsiadlowski2017}
{Podsiadlowski}, P. 2017, in Handbook of Supernovae, ed. A.~W. {Alsabti} \&
  P.~{Murdin}, 635, \dodoi{10.1007/978-3-319-21846-5\_123}

\bibitem[{{Podsiadlowski} {et~al.}(1992){Podsiadlowski}, {Joss}, \&
  {Hsu}}]{Podsiadlowski92}
{Podsiadlowski}, P., {Joss}, P.~C., \& {Hsu}, J.~J.~L. 1992, \apj, 391, 246,
  \dodoi{10.1086/171341}

\bibitem[{{Quirola-V{\'a}squez} {et~al.}(2019){Quirola-V{\'a}squez}, {Bauer},
  {Dwarkadas}, {Badenes}, {Brandt}, {Nymark}, \& {Walton}}]{QV2019}
{Quirola-V{\'a}squez}, J., {Bauer}, F.~E., {Dwarkadas}, V.~V., {et~al.} 2019,
  \mnras, 490, 4536, \dodoi{10.1093/mnras/stz2858}

\bibitem[{{Ramakrishnan} \& {Dwarkadas}(2021)}]{Ram2021}
{Ramakrishnan}, V., \& {Dwarkadas}, V.~V. 2021, Research Notes of the American
  Astronomical Society, 5, 191, \dodoi{10.3847/2515-5172/ac1e28}

\bibitem[{{Ross} \& {Dwarkadas}(2017)}]{Ross2017}
{Ross}, M., \& {Dwarkadas}, V.~V. 2017, \aj, 153, 246,
  \dodoi{10.3847/1538-3881/aa6d50}

\bibitem[{{Ryder} {et~al.}(1993){Ryder}, {Staveley-Smith}, {Dopita}, {Petre},
  {Colbert}, {Malin}, \& {Schlegel}}]{Ryder1993}
{Ryder}, S., {Staveley-Smith}, L., {Dopita}, M., {et~al.} 1993, \apj, 416, 167,
  \dodoi{10.1086/173223}

\bibitem[{{Salamanca} {et~al.}(1998){Salamanca}, {Cid-Fernandes},
  {Tenorio-Tagle}, {Telles}, {Terlevich}, \& {Munoz-Tunon}}]{Sala1998}
{Salamanca}, I., {Cid-Fernandes}, R., {Tenorio-Tagle}, G., {et~al.} 1998,
  \mnras, 300, L17, \dodoi{10.1046/j.1365-8711.1998.02093.x}

\bibitem[{{Sambruna} {et~al.}(2001){Sambruna}, {Brandt}, {Chartas}, {Netzer},
  {Kaspi}, {Garmire}, {Nousek}, \& {Weaver}}]{Sambruna2001}
{Sambruna}, R.~M., {Brandt}, W.~N., {Chartas}, G., {et~al.} 2001, \apjl, 546,
  L9, \dodoi{10.1086/318067}

\bibitem[{{Schlegel} {et~al.}(1998){Schlegel}, {Finkbeiner}, \&
  {Davis}}]{Schlegel1998}
{Schlegel}, D.~J., {Finkbeiner}, D.~P., \& {Davis}, M. 1998, \apj, 500, 525,
  \dodoi{10.1086/305772}

\bibitem[{{Siegel} {et~al.}(2021){Siegel}, {Dwarkadas}, {Frank}, \&
  {Burrows}}]{siegel21}
{Siegel}, J., {Dwarkadas}, V.~V., {Frank}, K.~A., \& {Burrows}, D.~N. 2021,
  \apj, 922, 67, \dodoi{10.3847/1538-4357/ac2305}

\bibitem[{{Smith}(2017)}]{Smith2017}
{Smith}, N. 2017, {Interacting Supernovae: Types IIn and Ibn}, ed. A.~W.
  {Alsabti} \& P.~{Murdin}, 403, \dodoi{10.1007/978-3-319-21846-5_38}

\bibitem[{{Soria} \& {Perna}(2008)}]{Soria2008}
{Soria}, R., \& {Perna}, R. 2008, \apj, 683, 767, \dodoi{10.1086/589995}

\bibitem[{{Summers} {et~al.}(2003){Summers}, {Stevens}, {Strickland}, \&
  {Heckman}}]{Summers2003}
{Summers}, L.~K., {Stevens}, I.~R., {Strickland}, D.~K., \& {Heckman}, T.~M.
  2003, \mnras, 342, 690, \dodoi{10.1046/j.1365-8711.2003.06590.x}

\bibitem[{{Taubenberger} {et~al.}(2015){Taubenberger}, {Elias-Rosa},
  {Kerzendorf}, {Hachinger}, {Spyromilio}, {Fransson}, {Kromer}, {Ruiter},
  {Seitenzahl}, {Benetti}, {Cappellaro}, {Pastorello}, {Turatto}, \&
  {Marchetti}}]{Taub2015}
{Taubenberger}, S., {Elias-Rosa}, N., {Kerzendorf}, W.~E., {et~al.} 2015,
  \mnras, 448, L48, \dodoi{10.1093/mnrasl/slu201}

\bibitem[{{Tully} {et~al.}(2009){Tully}, {Rizzi}, {Shaya}, {Courtois},
  {Makarov}, \& {Jacobs}}]{Tully2009}
{Tully}, R.~B., {Rizzi}, L., {Shaya}, E.~J., {et~al.} 2009, \aj, 138, 323,
  \dodoi{10.1088/0004-6256/138/2/323}

\bibitem[{{Tully} {et~al.}(2008){Tully}, {Shaya}, {Karachentsev}, {Courtois},
  {Kocevski}, {Rizzi}, \& {Peel}}]{Tully2008}
{Tully}, R.~B., {Shaya}, E.~J., {Karachentsev}, I.~D., {et~al.} 2008, \apj,
  676, 184, \dodoi{10.1086/527428}

\bibitem[{{Turatto} {et~al.}(1993){Turatto}, {Cappellaro}, {Danziger},
  {Benetti}, {Gouiffes}, \& {della Valle}}]{Turatto1993}
{Turatto}, M., {Cappellaro}, E., {Danziger}, I.~J., {et~al.} 1993, \mnras, 262,
  128, \dodoi{10.1093/mnras/262.1.128}

\bibitem[{{van Dokkum}(2001)}]{vanDokkum01}
{van Dokkum}, P.~G. 2001, \pasp, 113, 1420, \dodoi{10.1086/323894}

\bibitem[{{Vernet} {et~al.}(2011){Vernet}, {Dekker}, {D'Odorico}, {Kaper},
  {Kjaergaard}, {Hammer}, {Randich}, {Zerbi}, {Groot}, {Hjorth}, {Guinouard},
  {Navarro}, {Adolfse}, {Albers}, {Amans}, {Andersen}, {Andersen}, {Binetruy},
  {Bristow}, {Castillo}, {Chemla}, {Christensen}, {Conconi}, {Conzelmann},
  {Dam}, {de Caprio}, {de Ugarte Postigo}, {Delabre}, {di Marcantonio},
  {Downing}, {Elswijk}, {Finger}, {Fischer}, {Flores}, {Fran{\c{c}}ois},
  {Goldoni}, {Guglielmi}, {Haigron}, {Hanenburg}, {Hendriks}, {Horrobin},
  {Horville}, {Jessen}, {Kerber}, {Kern}, {Kiekebusch}, {Kleszcz}, {Klougart},
  {Kragt}, {Larsen}, {Lizon}, {Lucuix}, {Mainieri}, {Manuputy}, {Martayan},
  {Mason}, {Mazzoleni}, {Michaelsen}, {Modigliani}, {Moehler}, {M{\o}ller},
  {Norup S{\o}rensen}, {N{\o}rregaard}, {P{\'e}roux}, {Patat}, {Pena}, {Pragt},
  {Reinero}, {Rigal}, {Riva}, {Roelfsema}, {Royer}, {Sacco}, {Santin},
  {Schoenmaker}, {Spano}, {Sweers}, {Ter Horst}, {Tintori}, {Tromp}, {van
  Dael}, {van der Vliet}, {Venema}, {Vidali}, {Vinther}, {Vola}, {Winters},
  {Wistisen}, {Wulterkens}, \& {Zacchei}}]{Vernet2011}
{Vernet}, J., {Dekker}, H., {D'Odorico}, S., {et~al.} 2011, \aap, 536, A105,
  \dodoi{10.1051/0004-6361/201117752}

\bibitem[{{Weil} {et~al.}(2020{\natexlab{a}}){Weil}, {Fesen}, {Patnaude}, \&
  {Milisavljevic}}]{Weil2020a}
{Weil}, K.~E., {Fesen}, R.~A., {Patnaude}, D.~J., \& {Milisavljevic}, D.
  2020{\natexlab{a}}, arXiv e-prints, arXiv:2006.02496.
\newblock \doarXiv{2006.02496}

\bibitem[{{Weil} {et~al.}(2020{\natexlab{b}}){Weil}, {Fesen}, {Patnaude},
  {Raymond}, {Chevalier}, {Milisavljevic}, \& {Gerardy}}]{weil20b}
{Weil}, K.~E., {Fesen}, R.~A., {Patnaude}, D.~J., {et~al.} 2020{\natexlab{b}},
  \apj, 891, 116, \dodoi{10.3847/1538-4357/ab76bf}

\bibitem[{{Weiler} {et~al.}(2007){Weiler}, {Williams}, {Panagia}, {Stockdale},
  {Kelley}, {Sramek}, {Van Dyk}, \& {Marcaide}}]{Weiler2007}
{Weiler}, K.~W., {Williams}, C.~L., {Panagia}, N., {et~al.} 2007, \apj, 671,
  1959, \dodoi{10.1086/523258}

\bibitem[{{Winkler} {et~al.}(2017){Winkler}, {Blair}, \& {Long}}]{Winkler2017}
{Winkler}, P.~F., {Blair}, W.~P., \& {Long}, K.~S. 2017, \apj, 839, 83,
  \dodoi{10.3847/1538-4357/aa683d}

\bibitem[{{Woosley}(2010)}]{Woosley2010}
{Woosley}, S.~E. 2010, \apjl, 719, L204, \dodoi{10.1088/2041-8205/719/2/L204}

\bibitem[{{Xi} {et~al.}(2019){Xi}, {Gaetz}, {Plucinsky}, {Hughes}, \&
  {Patnaude}}]{xi19}
{Xi}, L., {Gaetz}, T.~J., {Plucinsky}, P.~P., {Hughes}, J.~P., \& {Patnaude},
  D.~J. 2019, \apj, 874, 14, \dodoi{10.3847/1538-4357/ab09ea}

\bibitem[{{Yamaguchi} {et~al.}(2014){Yamaguchi}, {Badenes}, {Petre}, {Nakano},
  {Castro}, {Enoto}, {Hiraga}, {Hughes}, {Maeda}, {Nobukawa}, {Safi-Harb},
  {Slane}, {Smith}, \& {Uchida}}]{Yamaguchi2014}
{Yamaguchi}, H., {Badenes}, C., {Petre}, R., {et~al.} 2014, \apjl, 785, L27,
  \dodoi{10.1088/2041-8205/785/2/L27}

\bibitem[{{Yamaguchi} {et~al.}(2018){Yamaguchi}, {Tanaka}, {Wik}, {Rho},
  {Bamba}, {Castro}, {Smith}, {Foster}, {Uchida}, {Petre}, \&
  {Williams}}]{Yamaguchi2018}
{Yamaguchi}, H., {Tanaka}, T., {Wik}, D.~R., {et~al.} 2018, \apjl, 868, L35,
  \dodoi{10.3847/2041-8213/aaf055}

\end{thebibliography}
\bibliographystyle{aasjournal}

\end{document}